\documentclass[aps,prb,reprint,twocolumn,superscriptaddress,showpacs]{revtex4-1}
 \pdfoutput=1
\newcommand{\be}{\begin{equation}}
\newcommand{\ee}{\end{equation}}
\newcommand{\bea}{\begin{eqnarray}}
\newcommand{\eea}{\end{eqnarray}}
\newcommand{\br}{{\bf r}}
\newcommand{\bJ}{{\bf j}}

\newcommand{\bj}{{\bf j}}

\newcommand{\bk}{{\bf k}}

\newcommand{\un}{\underline{n}}
 
\newcommand{\cV}{\mathcal V} 
\newcommand{\cN}{\mathcal N} 
\newcommand{\cC}{\mathcal C} 
\newcommand{\bq}{\mathbf q} 
\newcommand{\buk}{\underline{\mathbf k}}
 
\newcommand{\cq}{\mathfrak{q}}

\newcommand{\ualpha}{{\underline\alpha}}
\newcommand{\ubeta}{{\underline\beta}}

\newcommand{\blangle}{{\boldsymbol \langle\!\!\langle}}
\newcommand{\brangle}{{\boldsymbol \rangle\!\!\rangle}}

\newcommand{\dtP}{\partial_t {\bf P}}

\newcommand{\intbz}{\int\limits_\text{BZ}}
\newcommand{\iintbz}{\iint\limits_\text{BZ}}
\newcommand{\intbzdk}{\intbz d\bk}
\newcommand{\intbzdkpi}{\intbz\hspace{-0.15em}\frac{d\bk}{(2\pi)^d}}

\newcommand{\intbzdkpiprime}{\intbz\frac{d\bk'}{(2\pi)^d}}
\newcommand{\iintbzdkpi}{\iintbz\frac{d\bk}{(2\pi)^d}\frac{d\bk'}{(2\pi)^d}}
\newcommand{\intdkpi}{{\int}\frac{d\bk}{(2\pi)^d}}

\newcommand{\rhoC}{\rho}

\newcommand{\rhoD}{ \rho^\text{D}}

\newcommand{\rhoDnn}{{\rho^\text{D}_{nn}}}
\newcommand{\rhoDdotnprimen}{{\dot\rho^\text{D}_{n'n}}}
\newcommand{\rhoDnnprime}{{\rho^\text{D}_{nn'}}}
\newcommand{\rhoDnprimen}{{\rho^\text{D}_{n'n}}}
\newcommand{\rhoDnun}{{\rho^\text{D}_{n\un}}}
\newcommand{\rhoDunnprime}{{\rho^\text{D}_{\un n'}}}
\newcommand{\Trn}{\text{Tr}_n}

\newcommand{\vF}{v_\text{F}}

\newcommand{\mfBdef}{\mfT e^{\,\int\limits_{-\infty}^{t} \hspace{-0.2em}dt'\, \dot\bk\,\partial_\bk}}
\newcommand{\mfBdefinv}{\mfT e^{-\hspace{-0.3em}\int\limits_{-\infty}^{t} \hspace{-0.2em}dt'\, \dot\bk\,\partial_\bk}}
\newcommand{\eqt}{\,{=}\,}
\newcommand{\gt}{\,{>}\,}
\newcommand{\apt}{\,{\approx}\,}
\newcommand{\timest}{\,{\times}\,}

\newcommand{\pt}{\,{+}\,}

\usepackage{graphicx,color,amssymb,amsfonts}
\definecolor{red}{rgb}{1.0,0,0}
\definecolor{green}{rgb}{0,0.75,0}
\definecolor{pink}{rgb}{0,1,1}
\definecolor{blue}{rgb}{0,0,1}
\definecolor{orange}{rgb}{1,0.4,0}
\definecolor{darkgreen}{rgb}{0.0,0.5,0.0}

\usepackage{bm}
\usepackage{amsmath}
\usepackage{amssymb}
\usepackage{amsfonts}
\usepackage{upgreek}
\usepackage{mathbbol}
\usepackage{mathtools} 
\usepackage[ colorlinks,
    linkcolor={blue},
    citecolor={blue},
    urlcolor={blue}]{hyperref}
\usepackage{bbold} 
\usepackage[normalem]{ulem}
\usepackage{cleveref}

\DeclareMathAlphabet{\mathpzc}{OT1}{pzc}{m}{it}
\DeclareMathAlphabet\mathbfcal{OMS}{cmsy}{b}{n}

\newcommand{\ci}{{\mathfrak i}}

\newcommand{\bp}{{\bf p}}

\newcommand{\bR}{{\mathbf R}}

\newcommand{\bE}{{\bf E}}
\newcommand{\bB}{{\bf B}} 
\newcommand{\bA}{{\bf A}} 
\newcommand{\bP}{{\bf P}} 

\newcommand{\brho}{\boldsymbol{\rho}} 

\newcommand{\bfd}{\mathbf{d}}

\newcommand{\bfdtilde}{\tilde{\bfd}}
\newcommand{\mfB}{\mathfrak{B}} 
\newcommand{\mfT}{\mathfrak{T}}
	
\newcommand{\sbe}{{SBE}} 
\newcommand{\eom}{EoM}

\newcommand{\sh}{h^{(0)}}
\newcommand{\inh}{h^\text{in}}
\newcommand{\inrho}{\varrho^\text{in}}
\newcommand{\nB}{N_{\mathcal B}}
\newcommand{\tdHF}{time-dependent Hartree-Fock}
\newcommand{\vhf}{v_\text{HF}} 

\newcommand{\tepsilon}{\tilde{\epsilon}}

\newcommand{\bfcA}{\mathbfcal{A}}
\newcommand{\bfcAtilde}{\tilde{\bfcA}}

\newcommand{\refsemissionformula}{Golde2008,Golde2010,Vampa2014}
\newcommand{\dtPinter}{\dtP^\text{inter}(t)}
\newcommand{\dtPinterscalar}{\partial_tP^\text{inter}(t)}
\newcommand{\dtPinterx}{\partial_tP^\text{inter}_x(t)}
\newcommand{\jintratilde}{\tilde{\bj}^\text{intra}(t)}
\newcommand{\jintratildex}{\tilde{j}_x^\text{intra}(t)}

\newcommand{\jintrascalar}{j^\text{intra}(t)}

\newcommand{\jinter}{\bJ^\text{inter}(t)}
\newcommand{\jintertilde}{\tilde{\bJ}^\text{inter}(t)}
\newcommand{\jintertildex}{\tilde{j}^\text{inter}_x(t)}

\newcommand{\jinterx}{j^\text{inter}_x(t)}
\newcommand{\jintra}{\bJ^\text{intra}(t)}
\newcommand{\jintrax}{j_x^\text{intra}(t)}

\newcommand{\Jdamp}{\bJ^\text{deph}(t)}
\newcommand{\jdampx}{j^\text{deph}_x(t)}
\newcommand{\bomega}{\boldsymbol{\Omega}}
\newcommand{\sigmainter}{\sigma(t)}
\newcommand{\sigmamatel}{\sigma_{ij}(t)}
\newcommand{\aket}{\widetilde{|n\bk;t\brangle}}
\newcommand{\abra}{\widetilde{\blangle n\bk;t|}}
\newcommand{\aketprime}{\widetilde{|n'\bk;t\brangle}}

\newcommand{\aketu}{\widetilde{|\un\bk;t\brangle}}
\newcommand{\abrau}{\widetilde{\blangle \un\bk;t|}}
\newcommand{\ketcom}{|n\bk_t\brangle}
\newcommand{\ketcomprime}{|n'\bk_t\brangle}
\newcommand{\bracom}{\blangle n\bk_t|}
\newcommand{\rhoad}{\tilde{\varrho}}
\newcommand{\emikr}{e^{-i\bk\br}}

\newcommand{\eikrprime}{e^{i\bk\br'}}
\newcommand{\emikrR}{e^{-i\bk(\br+\bR)}}

\newcommand{\eikrRprime}{e^{i\bk(\br'+\bR')}}
\newcommand{\brR}{\br{+}\bR}
\newcommand{\brRprime}{\br'{+}\bR'}

\newcommand{\kt}{\bk_t}
\newcommand{\ktA}{\bk{-}\bA(t)}

\newcommand{\ketnkt}{|n\bk;t\brangle}
\newcommand{\ketnksubt}{|n\kt\brangle}
\newcommand{\ketnprimeksubt}{|n'\kt\brangle}
\newcommand{\ketunksubt}{|\un\kt\brangle}

\newcommand{\ketnk}{|n\bk\brangle}

\newcommand{\ketnkprime}{|n'\bk\brangle}
\newcommand{\brank}{\blangle n\bk|}
\newcommand{\brankprime}{\blangle n'\bk|}

\newcommand{\branksubt}{\blangle n\kt|}
\newcommand{\braunksubt}{\blangle \un\kt|}
\newcommand{\conti}{D}
\newcommand{\comoving}{C}
\newcommand{\SDK}{S}
\newcommand{\hilbert}{\mathcal{H}}
\newcommand{\emJW}[1]{\textit{#1}}

\newcommand{\cone}{\vF (k_y\sigma_x - k_x\sigma_y)}

\newcommand{\dnnprimek}{\mathbf{d}_{nn'}(\bk)}

\begin{document}

\title{{Semiconductor-Bloch-Equations Formalism: Derivation and Application to High-Harmonic Generation from Dirac Fermions}}
\author{Jan Wilhelm}\email{jan.wilhelm@physik.uni-regensburg.de}
\author{Patrick Gr\"ossing}
\author{Adrian Seith} 
\affiliation{Institute  of  Theoretical  Physics,  University  of  Regensburg, Universit\"atsstra\ss e  31,  D-93053  Regensburg,  Germany}
\author{Jack Crewse}
\affiliation{Institute  of  Theoretical  Physics,  University  of  Regensburg, Universit\"atsstra\ss e  31,  D-93053  Regensburg,  Germany}
\affiliation{Department of Physics, Missouri University of Science \& Technology, Rolla, Missouri 65409, USA}
\author{\mbox{Maximilian Nitsch}}
\affiliation{Institute  of  Theoretical  Physics,  University  of  Regensburg, Universit\"atsstra\ss e  31,  D-93053  Regensburg,  Germany}
 \author{Leonard Weigl}
 \author{Christoph Schmid}
\affiliation{Institute of Experimental and Applied Physics,  University  of  Regensburg, Universit\"atsstra\ss e   31,  D-93053  Regensburg,  Germany}
\author{Ferdinand Evers}\email{ferdinand.evers@physik.uni-regensburg.de}
\affiliation{Institute  of  Theoretical  Physics,  University  of  Regensburg, Universit\"atsstra\ss e  31,  D-93053  Regensburg,  Germany}

\date{\today}
\begin{abstract}

We rederive the semiconductor Bloch equations emphasizing the close link to  the  Berry  connection. 
Our rigorous derivation reveals the existence of two further contributions to the current, in addition to the frequently considered  intraband and polarization-related interband terms. 
 The extra contributions become sizable in situations with strong dephasing or when the dipole-matrix elements are strongly wave-number dependent. 
We apply the formalism to high-harmonic generation for a Dirac metal. 
The extra terms add to the  frequency-dependent emission intensity (high-harmonic spectrum) significantly at certain frequencies changing the total signal up to a factor of 10.

\end{abstract}

\maketitle


\section{Introduction}

The advancement of time-resolved spectroscopy seen in recent years facilitated the  study  of dynamical processes on sub-cycle time scales. 
Interesting effects that arise along the way in metals and semiconductors  include the generation of high-harmonics (HHG) by the lightwave-driven charge carriers~\cite{Ghimire2011,Schubert2014,Vampa2015,Hohenleutner2015,Ndabashimiye2016,Garg2016,Ghimire2019}, 
subcycle control of charge transport in nanostructures~\cite{Rybka2016},  and atomic-resolution ultrafast microscopy~\cite{Cocker2016}.
Since high-harmonics are very sensitive to acceleration processes that the charge carriers are subjected to, HHG can be used for monitoring dynamical processes. Promising applications for band structure reconstruction~\cite{Vampa2015b,Tancogne2017} and for observing dynamical Bloch oscillations~\cite{Schubert2014,Luu2015} and  Berry phase effects~\cite{Luu2018,Liu2017,Silva2019} 
have been reported. 

An established theoretical framework to describe the dynamics of quantum systems is the density-matrix formalism 
that is known as semiconductor Bloch equations (\sbe) in the context of crystalline solids.\cite{SchmittRink1988, Lindberg1988,Haug2009,Kira2011,Schaefer2002,Haug2008}
It is exact, in principle, but in many applications of SBE to strong field dynamics, such as HHG, dynamical contributions from Coulomb interactions are neglected while band-structure effects are properly kept.
This approximation has proven to be  useful in numerous applications including, in particular,  HHG in various model systems~\cite{Kruchinin2013,Vampa2014,Hawkins2013,Luu2016,Tamaya2016,Silva2019,Fu2020} and materials~\cite{AlNaib2014,Wismer2016,Dimitrovski2017,Jiang2018,Zhang2019}.

In the first part of the article, we present a rederivation of the main equations of motion (\eom) 
for the density matrix $\rho$ and the physical observables in the framework of {\sbe}. 
The particular perspective we here offer emphasizes the close relation between {\sbe} and the Berry connection.

Second, we present a rigorous derivation of the relation between $\rho$ and the longitudinal current $\bj(t)$. In the absence of phenomenological dephasing, such a relation has already been derived in earlier works by Sipe and coworkers.\cite{Aversa1995, Sipe2000} These works embark on a decomposition of the position operator into intra- and interband contributions and focus on the polarization as the central observable. The perspective we here advocate is based on the current density; the final splitting of observables into intra- and interband contributions then is a result of the calculation.
Further, we consider the {\sbe} including a phenomenological dephasing. As it turns out, this gives an extra contribution to the current density, which gets sizeable for Dirac fermions and has not been discussed before.

Our exact result contains several terms that are not
accounted for in earlier work~\cite{\refsemissionformula}.
The  consequences will be discussed for the example of Dirac fermions: we find  qualitative agreement with respect to the high-harmonic spectrum between the exact and the approximated expression. Quantitative discrepancies appear, however, which can exceed an order of magnitude.
The manuscript is organized as follows:  Sec.~\ref{EoMbasis} focuses on the {\eom} for the density matrix, $\rho(t)$, with emphasis on {\sbe} and the Berry connection. 
In Sec.~\ref{sec:observables}, we relate $\rho(t)$  to the time-dependent longitudinal current density and  the frequency-dependent emission intensity that underlies the HHG. 
An application to HHG in metallic films with Dirac-like spectrum is presented in Sec.~\ref{sec5}.

\section{Derivation of equations of motion for the density matrix} \label{EoMbasis}

The power of the density matrix formalism is in its simplicity. In principle, it allows for the 
propagation of observables in a genuine many-body theory keeping the effective Hilbert space on the single-particle size. It thus can be intrinsically more efficient than wavefunction correlation theory~\cite{Vidal2007,Imai2020,White1992,Weiner2019}. 
Conceptually similar are Green's function based approaches, such as $GW$ + Bethe-Salpeter~\cite{onida2002,Golze2019,Attaccalite2011,Attaccalite2017,Attaccalite2019,Golez2019}. They keep an additional dynamical degree of freedom, however, and therefore tend to be computationally more expensive. An affordable alternative to density-matrix based approaches is the 
time-dependent density functional theory~\cite{Runge1984,Provorse2016,Tancogne2020,Tancogne2017b,Tancogne2018,Tancogne2018a,Mrudul2020,Dewhurst2020,Wang2019,Noda2019,Andermatt2018,Otobe2009,Otobe2016,Hansen2017,Bauer2018,Jurss2019,Pemmaraju2018}.
It has the advantage that implementations are available that can treat inhomogeneous systems of considerable size; progress towards including spatially varying electric fields has been made only recently.~\cite{Floss2018}
We nevertheless here employ the density-matrix formalism, because it allows us to incorporate phenomenological damping terms that describe effects of dissipative environments. As it turns out, such terms are needed when comparing computational results with experimental data.

In this section, we  recall the derivation of the \sbe; we carefully define the mathematical objects entering later applications. 
We will derive general equations within the framework of Hartree-Fock theory. Later applications will be given for non-interacting electrons disregarding all correlation effects, such as 
 excitons~\cite{Wang2005,Merkl2019,Ovesen2019,Verdenhalven2013}, also phonons~\cite{Giustino2017,Cosacchi2018} and the quantization of electromagnetic fields~\cite{Flick2017,Hoffmann2019,Gombkoto2020,Rokaj2019}. As it turns out, for the qualitative description of many experimental findings, the non-interacting theory is a useful first step.


\subsection{Equation of motion} 

Consider the fermionic, second quantized many-body Hamiltonian 
\be 
\hat H = \sum_{\alpha\beta} \sh_{\alpha\beta}c^\dagger_{\alpha}c_\beta
+ \frac{1}{2} \sum_{\alpha,\beta,\gamma,\delta} U_{\alpha\beta\gamma\delta} c^\dagger_{\alpha}c^\dagger_{\beta}
c_\gamma c_\delta
\label{e1} 
\ee
with $\sh$ denoting a generic single-particle Hamiltonian 
represented in a stationary basis 
of a number of $\nB$ single-particle states  $|\phi^{(0)}_\alpha\rangle$.  
In the presence of time-dependent perturbations, such as external electric or magnetic fields, this component of $\hat H$ becomes time dependent, $\sh(t)$. 
The density matrix, $\rho$,  
is defined by the matrix elements 
\be
\rho_{\alpha\beta}(t) \coloneqq \langle \Psi(t)| c^\dagger_\beta c_\alpha|\Psi(t)\rangle. 
\label{e2} 
\ee
It describes selected aspects of a time evolving many-body state~$\Psi(t)$ that enter physical observables, e.g., the particle density.

The time-evolution of $\rho(t)$ derives 
directly from the definitions~\eqref{e1}, \eqref{e2} and the Schr\"odinger equation; in a basis-free representation the 
resulting {\eom} takes the form~\cite{Ring2004} 
\bea
\label{e3} 
\ci \dot \rho &=& [h(t), \rho] + \ci \left.\frac{\partial \rho}{\partial t}\right|_\text{coll}(t)\\
h(t) &\coloneqq& \sh(t) + \vhf(t).
\label{e4} 
\eea
While the commutator in \eqref{e3} 
accounts for the (effective) single-particle dynamics, the collision term $\left.\frac{\partial \rho}{\partial t}\right|_\text{coll}(t) $ includes genuine two-body effects. Systematic expansions have been proposed to deal with it approximately, however, at the expense of a considerable numerical effort.~\cite{Axt1994,Axt1996}

We here consider {\tdHF} theory, where the  collision term is dropped and the time evolution of $\rho$ remains unitary. In order to mimic the (non-unitary) effects of collisions, the collision term can be approximated on a heuristic level by replacing it with phenomenological damping terms~\cite{Schaefer2002}. For recent discussions on the strength and physical content of dephasing and damping in the context of semiconductor Bloch equations, we refer to Refs.~\onlinecite{Floss2018} and~\onlinecite{Kilen2020}. 
The mean-field interaction, $\vhf$, 
can be understood as a 
known~\cite{Ring2004} 
linear functional of the density matrix $\vhf(t)\coloneqq {\mathcal F}[\rho]$, 
\begin{align}
&{\mathcal F}_{\alpha\beta}(t)
\coloneqq 
\sum_{\ualpha\ubeta}
(U_{\alpha\ualpha\beta\ubeta}
- U_{\alpha\ualpha\ubeta\beta})
\ \rho_{\ubeta\ualpha}(t), 
 \label{e5} 
\end{align}
where the matrix elements $\rho_{\alpha\beta}(t)$ are the representation of $\rho(t)$ in the stationary basis $|\phi_\alpha\rangle$: 
$\rho_{\alpha\beta}(t)=\langle \phi_\alpha|\rho(t)|\phi_\beta\rangle$. 
The functional \eqref{e5} together with \eqref{e3} gives a closed set of equations for the dynamics of  $\rho(t)$. 
Exchange-correlation functionals alternative to 
Eq.~\eqref{e5} have been explored in the spirit of (time-dependent) density functional theory.~\cite{Floss2019} 

\subsection{The adiabatic basis} 

We define an adiabatic basis~\cite{Xiao2010}
$|\alpha;t\rangle$ 
by the simultaneous,
orthonormalized 
eigenstates of 
$h(t)$
\begin{align}
h(t)|\alpha;t\rangle = \tepsilon_\alpha(t)|\alpha;t\rangle. \label{e6} 
\end{align}
In this basis, the commutator dynamics \eqref{e3} takes a simple form.
Notice that \eqref{e6} defines the basis at time~$t$ only up to a phase factor. Therefore, two basis sets at neighboring times $t$ and $t+dt$, $|\alpha;t\rangle$ and~$|\alpha;t+dt\rangle$,  can differ, in principle, by an 
arbitrary phase factor so that the motion 
of matrix elements given in the adiabatic frame is
not yet uniquely defined. We conclude that the time 
evolution of the phase-factor needs to be imposed by an extra condition that complements \eqref{e6} but is not part of~\eqref{e6}.

In order to formulate this condition we adopt the attitude that $|\alpha;t\rangle$ and $|\alpha;t+dt\rangle$ should be smoothly connected in a manner as it would be implied by perturbation theory; we thus stipulate 
\be
\partial_t|\alpha;t\rangle \coloneqq \sum_{\beta\neq \alpha} |\beta;t\rangle \,\frac{\langle \beta;t|
\dot h(t)|\alpha;t\rangle}{\tepsilon_\alpha(t) - \tepsilon_\beta(t)}\,,
\label{e6a}
\ee 
which implies $\langle\alpha;t|\partial_t |\alpha;t\rangle = 0$. 
The time evolution~\eqref{e6a} starts at $t\rightarrow-\infty$ with initial eigenstates
 $|\alpha(-\infty)\rangle \coloneqq |\phi_\alpha\rangle$ 
that are defined as
\begin{align}
    \inh |\phi_\alpha\rangle = \epsilon_\alpha|\phi_\alpha\rangle\,,\hspace{2em}\inh\coloneqq\underset{t\rightarrow-\infty}{\lim}h(t)\,.\label{e9a}
\end{align}
To further connect  the time evolution~\eqref{e6a} to other definitions in the literature~\cite{Xiao2010}, we specify to a situation where $h(t)$ is implicitly time dependent, because it contains a set of parameters $\bR(t)$ that are time dependent, $h[\bR(t)]$. These parameters could be, e.g., external electric or magnetic fields, $\bE(t)$ and $\bB(t)$, but in the case of self-consistent field theories also the matrix elements of $\rho$ themselves. We thus have
\be
\partial_t|\alpha;t\rangle = \dot \bR \sum_{\beta\neq \alpha} |\beta;t\rangle \frac{\langle \beta;t|
\frac{\partial h}{\partial \bR}|\alpha;t\rangle}{\tepsilon_\alpha(t) - \tepsilon_\beta(t)}. \label{e9}
\ee 
Suppressing the time-dependencies in our notation, the matrix element can be evaluated by observing that 
\bea
\partial_\bR \langle \beta|h|\alpha\rangle &=& \langle \partial_\bR\beta|h|\alpha\rangle + \langle\beta|\partial_\bR h|\alpha\rangle + \langle\beta|h|\partial_\bR\alpha\rangle\nonumber\\
&=& \tepsilon_\alpha\langle\partial_\bR\beta|\alpha\rangle + \tepsilon_\beta\langle \beta|\partial_\bR\alpha\rangle + \langle \beta|\partial_\bR h|\alpha\rangle \nonumber\,.
\eea 
Since $\langle \beta|\partial_\bR\alpha\rangle = - \langle \partial_\bR \beta|\alpha\rangle$,
we have 
\be
\langle \beta|\partial_\bR h|\alpha\rangle = (\tepsilon_\alpha - \tepsilon_\beta)\langle\beta|\partial_\bR \alpha\rangle + \delta_{\alpha\beta}\partial_\bR \tepsilon_\alpha.
\ee 
When inserting this relation into \eqref{e9} we arrive at the result
\be
\partial_t |\alpha;t\rangle \coloneqq 
\dot \bR(t) \sum_{\beta\neq \alpha} |\beta;t\rangle\langle\beta;t|\partial_\bR|\alpha;t\rangle. 
\label{e8} 
\ee
We adopt the formulation of dynamics in the adiabatic basis as in Eq.~\eqref{e8} as our preferred one. 
It reveals the close connection to 
differential geometry, because it implies 
\be 
\dot\bR(t) \langle \alpha;t|\partial_\bR|\alpha;t\rangle 
= 0
\label{e12} 
\ee
that we have obtained from  $\langle\alpha;t|\partial_t|\alpha;t\rangle=0$, see note below~\eqref{e6a}.
Relation \eqref{e12} is well known as the {\it condition of parallel transport}~\cite{Xiao2010};   it is a result of the specific way to define the phase
evolution of wavefunctions~$|\alpha;t\rangle$ during time  by imposing \eqref{e6a}.
Eq. \eqref{e12} implies that the motion of the adiabatic frame is such that the
{\em Berry connection}~\cite{Berry1984}
\be 
\bfcA_\alpha[\bR]\coloneqq \langle \alpha;t|\ci\partial_\bR|\alpha;t\rangle
 \ee 
 remains perpendicular to the "velocity"
 of
 each state $|\alpha;t\rangle$. 
 \be 
 \dot \bR(t) \cdot \bfcA_\alpha[\bR]=0\,. 
\label{e14} 
\ee 
We further illustrate the meaning of \eqref{e8} 
discussing the example of Bloch electrons in homogeneous electric field. 

\subparagraph*{Bloch electrons in homogeneous $\bE(t)$.} We consider charged free fermions, so $\vhf{\to}0$
and $h{\to}\sh$. They are embedded in a crystal lattice, so 
the eigenstates of the stationary single particle Hamiltonian  (without electric field, $\bE(t)=0$) are Bloch-states $|n\bk\rangle$, which implies  $|\alpha\rangle {\to} |n\bk\rangle$.
Recalling Bloch's theorem, we have a factorization of the eigenstates
\be
\langle \br|n\bk\rangle = \frac{1}{\sqrt{\cN}}\, e^{\ci\bk\br} \blangle \br|n\bk\brangle  \label{e15}
\ee
with eigenvalues $\epsilon_n(\bk)$; 
here, $\cN$ denotes the number of unit cells and the matrix element on the rhs represent the lattice-periodic content of the Bloch state,  $u_{n\bk}(\br)\coloneqq\blangle\br|n\bk\brangle$  in a traditional notation~\cite{Ashcroft1976}. 
The double angular brackets indicate that the normalization volume for $u_{n\bk}$ is the unit cell, see Appendix~\ref{a0} for more details on our notation. 
Formally, the states $|n\bk\brangle$ are solutions of the  eigenvalue problem 
\begin{align}
{\inh (\bk) \,|n\bk\brangle = 
\epsilon_{n}(\bk) \,|n\bk\brangle}
\label{e16}
\end{align}
with 
\begin{align} 
h^\text{in}(\bk) \coloneqq 
\sum_{n}|n\bk\brangle \epsilon_{n}(\bk)
\blangle n\bk|
\label{e17} 
\end{align} 
see Eq.~\eqref{b12} in Appendix~\ref{a0}. 
 Summarizing, the stationary Bloch-Hamiltonian reads 
\begin{align}
    \hat H = \sum_{mm'} \intbzdkpi\
    h^\text{in}_{mm'}(\bk)
    \ c^\dagger_{m}(\bk)c_{m'}(\bk) 
\end{align}
where $h_{mm'}(\bk)\coloneqq\blangle m|h^\text{in}(\bk)|m'\brangle$ and the states $|m\brangle$ denote a generic basis in the subspace of the degrees of freedom of the unit cell (bands) that may or may not be chosen to depend on $\bk$. 

As a time-dependent perturbation acting on 
fermions of charge $\cq$, we introduce a homogeneous electric field $\bE(t)$  that evolves from zero, i.e.~$\underset{t\rightarrow-\infty}{\lim}\bE(t) = 0$. Its effect is discussed conveniently in the Coulomb gauge~\footnote{The Coulomb gauge is defined as $\nabla\cdot\bA\eqt 0$.~\cite{Jackson2009}
In our application, source terms for generating electric fields are absent, i.e. $\Delta\Phi\eqt0$ and $\text{div} \,{\bE}=0$. In principle, a gauge-degree of freedom is left in this case. It implies possibilities for alternative representations, e.g., $\Phi(\br,t) {=} -\br \bE(t)$ with the longitudinal component of $\bA$ being independent of time ('length gauge') or $\bA(t) {=}{-}\cq \int_{-\infty}^{t} dt'\, \bE(t')$ with $\Phi$ being independent of position ('velocity gauge');~\cite{Foeldi2017} 
 evaluating the expression $\bE = -\nabla \Phi - \dot\bA/\cq$ in either gauge, the same electric field is reproduced. 
 The representation of electric potential via the length gauge frequently occurs in the context of dipole expansions. In the literature, the velocity gauge and the Coulomb gauge are often identified with each other; for further discussion see Ref.~\onlinecite{Jackson2009}\label{foot:gauge}
}
\begin{align}
\cq \bE(t) = -\dot \bA(t) \label{e18a},
\end{align}
where a  
factor $\cq/c$ was absorbed in the definition of~$\bA$. 
As compared to the alternative gradient represention, $\bE(t) {=} -\nabla \phi(\br,t)$,  the Coulomb gauge offers the advantage that it does not break translational invariance for homogeneous electric fields; therefore, it is particularly convenient for treating Bloch electrons. Using minimal coupling, we have~\cite{Kira2011,Xiao2010,Altland2010}
\be
h(\bk; t) \coloneqq \inh(\bk_t)\,,
\hspace{1.2em}
\bk_t = \bk - \bA(t)
\label{e17a}
\ee
and correspondingly 
\begin{align}
\inh (\bk_t) \,\aket  = 
\tepsilon_n(\bk;t) \  \aket 
\label{e18} 
\end{align}
with the analogies $\bR(t) {\to} \bA(t)$ and $|\alpha;t\rangle {\to} \aket$. 
  Due to minimal coupling~\eqref{e17a}, the eigenvalues are  given by
\be
\tepsilon_n(\bk; t) = \epsilon_{n}(\bk_t)\,. \label{e19a}
\ee
The tilde on~$\aket$ emphasizes the adiabatic time evolution from \eqref{e6a}/\eqref{e8},~\footnote{
Note that due to translational invariance, 
 only diagonal matrix elements 
with $\bk{=}\bk'$ appear in \eqref{e20} and an additional sum~$\sum_{\bk'}$ is absent.
We illustrate in Appendix~\ref{a1}, \eqref{e54}  that contributions from off-diagonals $\bk{\neq}\bk'$ vanish.}
\be
\partial_t \aket = 
-\cq \bE(t) \sum_{\un\neq n} \aketu
\abrau
\frac{\partial}{\partial \bA}
\aket\, \label{e20}
\ee
such that the condition of parallel transport \eqref{e14} in the adiabatic basis~$\aket$ is satisfied,
\begin{align}
\bE(t)\ \abra
\frac{\partial}{\partial \bA}
\aket 
= 0\,.
\end{align}
We note that the matrix elements used for the time evolution \eqref{e20} are  
\begin{align}
\abra\ci\partial_t\aketprime &=
 -\bE(t)\cq \abra\ci\frac{\partial}{\partial \bA}\aketprime \nonumber
\\
 &=
\bE(t) \bfdtilde_{nn'}(\bk;t) \label{e25}
\end{align}
introducing the dipole matrix element 
\be
\bfdtilde_{nn'}(\bk;t) \coloneqq - \cq \abra\ci\frac{\partial}{\partial \bA}\aketprime, \label{e22}
\ee
with diagonal elements 
\be 
\bfcAtilde_n(\bk;t) \coloneqq -\cq \abra\ci\frac{\partial}{\partial \bA}\aket\label{e23}
\ee 
known as the {\it Berry connection}.
We arrive at a compact notation for the condition of parallel transport, 
\begin{align}
\bE(t) \cdot \bfcAtilde_n(\bk;t) = 0. \label{e24}
\end{align}


\subsection{{\eom} for the density matrix in adiabatic basis} 
In the adiabatic basis defined in~\eqref{e6} and \eqref{e6a}, the {\eom}~\eqref{e3}  takes the form (in the absence of collisions)
\begin{align}
\ci \langle \alpha;t|\dot \rho|\beta;t\rangle &=  \tepsilon_{\alpha\beta}(t)
\varrho_{\alpha\beta}(t) \,,
\label{e19} 
\end{align}
where we define $\tepsilon_{\alpha\beta}{(t)} = \tepsilon_\alpha{(t)} {-} \tepsilon_\beta{(t)}$ and 
\[
\varrho_{\alpha\beta}(t) \coloneqq \langle\alpha;t|\rho(t)|\beta;t\rangle\,.
\]
To arrive at a closed set of equations for the matrix elements of $\rho$ in the adiabatic frame, we need to reformulate \eqref{e19} so time-derivatives of matrix elements of $\rho$ appear - rather than matrix elements of $\dot \rho$. To arrive at such an {\eom} for the matrix elements, we will employ the relation
\be
\ci \frac{d}{dt}
\langle \alpha;t|\rho|\beta;t\rangle=  
\tepsilon_{\alpha\beta}(t)
\varrho_{\alpha\beta} \nonumber\\
+ \ci \langle   \dot\alpha|\rho|\beta\rangle+ 
\ci \langle \alpha|\rho| \dot\beta\rangle
\nonumber 
\ee 
where \eqref{e19} has been used; on the rhs the time variable has been suppressed and a short-hand notation $\partial_t|\alpha;t\rangle {=} |\dot \alpha;t\rangle$ was introduced. Inserting the resolution of the identity,
$1=\sum_{\ualpha} |\ualpha\rangle\langle\ualpha|$,
we find 
\be
\left( \ci \frac{d}{dt}
- \tepsilon_{\alpha\beta}(t)\right)
\varrho_{\alpha\beta}=\ci   
\sum_{\ualpha} \langle  \dot\alpha|\ualpha\rangle \varrho_{\ualpha\beta} + 
\varrho_{\alpha\ualpha} \langle\ualpha|\dot\beta\rangle.
\label{e17c} 
\ee 
With  $  \langle   \alpha|\dot\ualpha\rangle =-\langle \dot  \alpha|\ualpha\rangle $ and Eq. \eqref{e8}, we conclude   
\be
\left( \ci \frac{d}{dt}
{-} \tepsilon_{\alpha\beta}(t)\right)
\varrho_{\alpha\beta}= \dot\bR(t)
\sum_{\ualpha} 
\varrho_{\alpha\ualpha} \langle\ualpha|\ci\frac{\partial\beta}{\partial\bR}\rangle-\langle \alpha|\ci\frac{\partial\ualpha}{\partial\bR}\rangle 
\varrho_{\ualpha\beta} 
\label{e10} 
\ee 
arriving at the explicit form of the general {\eom} in the adiabatic frame. 
%

\subparagraph*{Semiconductor Bloch equations.} 
In the presence of a crystal 
symmetry (and in the absence of mean-field interactions) the equation of motion of the density operator, Eq.~\eqref{e4},  takes a block-diagonal form 
\begin{align}
    \ci \dot \rho(\bk) 
    = [h(\bk;t),\rho(\bk)]\label{e30a}
\end{align}
where each block has a common~$\bk$-vector and, analogous to Eq. \eqref{e17}, $h(\bk;t)$ and $\rho(\bk)$ are matrices that act within the Hilbert space of the unit cell ("bands"). The matrices $h(\bk;t)$ and $\rho(\bk)$ are defined via their matrix elements: 
\begin{align}
    h_{nn'}(\bk;t) = \langle n\bk|h(t)|n'\bk\rangle
    = \blangle n\bk|h(\bk;t)|n'\bk\brangle 
    \label{e31a} 
\end{align}
and similarly for $\rho_{nn'}(\bk)$, 
see Appendix~\ref{a0} for further details. 
Electric fields are readily treated in the  Coulomb-gauge: 
$h(\bk;t)=\inh(\bk{-}\bA(t))$.
The stationary basis used 
in \eqref{e31a} can be rotated into the adiabatic Bloch states from \eqref{e15}-\eqref{e23} with the analogies $\bR(t)\,{\to}\, \bA(t)$ and $|\alpha;t\rangle\, {\to}\, \aket$.  The results of the previous section then translate into  
\begin{align} 
\begin{split}
\big( &\ci \frac{d}{dt}\
{-} \epsilon_{nn'}(\bk_t)\big)
\rhoad_{nn'}(\bk;t) =\\[0.3em]
&\bE(t) 
\sum_{\un} 
\rhoad_{n\un}(\bk;t)\bfdtilde_{\un n'}(\bk;t)
-\,\bfdtilde_{n\un}(\bk;t)\rhoad_{\un n'}(\bk;t)
\end{split}
\label{e11}
\end{align} 
with the density matrix~$\rhoad_{nn'}(\bk;t)$ in the adiabatic basis  
and defining
\[
\epsilon_{nn'}(\bk_t) \coloneqq \epsilon_n(\bk_t) - \epsilon_{n'}(\bk_t)\,.
\]
Eqs.~\eqref{e11} are known as the semiconductor Bloch equations (\sbe).~\cite{Schaefer2002,Kira2011,Haug2009} 
They have been derived here emphasizing a geometric perspective.  
Note that due to translational invariance, 
in \eqref{e11} only diagonal matrix elements of~$\rhoad$ with a single $\bk$-point  
appear, see Appendix~\ref{a1} for details.  
Another remarkable property of Eq. \eqref{e11} is that matrix elements 
$\rhoad_{nn'}(\bk;t)$ taken at different wavevectors $\bk$ 
do not couple due to translational invariance of $\bA(t)$; terms involving gradients $\partial_\bk$ 
are absent in \eqref{e11}, 
which otherwise appear; see Appendix \ref{a1} for further details.

\subsection{Co-moving basis and EoM for its density matrix}
We categorize the basis sets introduced before by considering a mapping $f{:}\,(\mathbb{N},\text{1.\,BZ},\mathbb{R})\to\hilbert$, where $\hilbert$ is the Hilbert space of Bloch states. 
We regard~$\ketnkt$ as such a function~$f$ with variables~$n$,~$\bk$ and~$t$, that, when evaluated for a given~$n$, $\bk$ and $t$, returns a state in $\hilbert$.
All of these functions are collected in the set
\begin{align*}
    &F \coloneqq
    \Big\{\ketnkt{:}\,(\mathbb{N},\text{1.\,BZ},\mathbb{R})\to\hilbert     \Big \}\,.
\end{align*}
%
We define the set of \emJW{instantaneous} functions~$I$   containing every function~$\ketnkt$ that is an eigenstate of~$h(\bk;t)$ for each instantaneous $(\bk,t)$ pair,
\begin{align*}
    I \coloneqq \Big\{ \ketnkt\in F : 
    h(\bk;t) \, \ketnkt
    \overset{\text{\eqref{e18}}}{=} 
\epsilon_n(\bk;t)   \,\ketnkt     \Big\} \,.
\end{align*}
Next, we define a subset of $I$ that has the special property that the phase factors evolve smoothly in time, i.e., the functions are differentiable in time,
\begin{align*}
    \conti \coloneqq \Big\{ \ketnkt \in I :\  & \text{$\ketnkt$  differentiable  in $t$}   \Big\} \,.
\end{align*}
In the same spirit, we define the \emJW{adiabatic} subset of functions that additionally fulfill the adiabatic time evolution~\eqref{e20},
\begin{align*}
    A \coloneqq \Big\{ & \aket \in D :  \\ &\partial_t \aket \overset{\text{\eqref{e20}}}{=}
-\cq {\bE(t)} \sum_{\un\neq n} \aketu
\abrau
\frac{\partial}{\partial \bA}
\aket  \Big\}\,. 
\end{align*}
We further define the set~$\SDK$ of stationary (i.e.~time-independent), differentiable-in-$\bk$ functions for a stationary basis $\ketnk{:}\,(\mathbb{N},\text{1.\,BZ})\,{\to}\,\hilbert$,
\begin{align*}
    \SDK\coloneqq \Big\{\ketnk:  
\inh&(\bk) |n\bk\brangle \overset{\text{\eqref{e16}}}{=} 
\epsilon_{n}(\bk) |n\bk\brangle
\\[-0.3em]
& 
   \text{ and $\ketnk$ differentiable  in $\bk$}   \Big\}\,.
\end{align*}
\textbf{Bloch electrons in homogeneous electric field.} For the dynamics of Bloch electrons in a homogeneous electric field, we have $h(\bk;t)=\inh(\bk{-}\bA(t))$.
It is convenient to introduce a set of \emJW{co-moving} functions as
\begin{align}
\begin{split}
    \comoving \coloneqq  
    \Big\{ \ketnkt \in D\, {:}\ &\text{there }\text{is a } \ketnk\in \SDK 
    \\[-0.3em] & \text{such that } \ketnkt = \ketcom \Big\}\,,
    \end{split}\label{e29a}
\end{align}
using the definition $\bk_t{=}\bk{-}\bA(t)$ from~\eqref{e17a}.
The co-moving set forms a basis that is also known as {\it Houston basis}~\cite{Houston1940} in the literature.
We mention that a co-moving function~$\ketnkt\,{\in}\,\comoving$ is an eigenstate of~$h(\bk;t)$ with eigenvalue~$\epsilon_n(\bk_t)$, see \eqref{e16}. 
In general, a co-moving function is not adiabatic,
\[
\comoving \not\subset A\,,
\]
that means, the condition of parallel transport, Eq.~\eqref{e24}, is violated by a general co-moving function.
The only degree of freedom that distinguishes between an adiabatic function~$\aket\in A$ and a co-moving function $\ketcom$ is a differentiable phase~\cite{Xiao2010}~$\gamma_n(\bk_t,t)$ such that
\begin{align}
\aket = \exp(i\gamma_n(\bk_t,t))\,\ketcom\,.\label{e30}
\end{align}
%

%
The dipole moment and the Berry connection from \eqref{e22} and \eqref{e23}
when expressed  in the co-moving basis
\eqref{e29a}, $\ketnkt{=}\ketcom$, turn into familiar expressions~\cite{Berry1984,Schaefer2002,Kira2011,Haug2009,Li2019}
\begin{align}
\bfd_{nn'}(\bk_t) &=  \cq \bracom\ci\partial_{\bk_t}\ketcomprime\,,\label{e32}
\\[0.5em]
\bfcA_n(\bk_t) &=  \cq \bracom\ci\partial_{\bk_t}\ketcom\,.\label{e33b}
\end{align}
For deriving an equation of motion for the density matrix in the co-moving basis~$\ketcom$, we proceed similarly as for deriving Eq.~\eqref{e11}:
In Eq.~\eqref{e10}, the substitutions $\bR(t)\,{\to}\, \bA(t)$ and $|\alpha;t\rangle\, {\to}\, \ketcom$ 
lead to the familiar form of the {\sbe} in the co-moving basis as~\cite{Schaefer2002,Kira2011,Haug2009,Li2019} 
\begin{align} 
\begin{split}
\Big( &\ci \partial_t
-\epsilon_{nn'}(\bk_t)\Big)
\varrho_{nn'}(\bk;t) =\\[0.3em]
&\bE(t) 
\sum_{\un} 
\varrho_{n\un}(\bk;t)\bfd_{\un n'}(\bk_t)
-\,\bfd_{n\un}(\bk_t)\varrho_{\un n'}(\bk;t)\,.
\end{split}
\label{e33a}
\end{align} 
The co-moving basis is our preferred basis for numerical calculations since dipoles and Berry connections, \eqref{e32} and \eqref{e33b}, are easy to compute.
In this representation, the {\sbe} constitute an $N_\text{b}$-level model, where $N_\text{b}$ represents the number of bands. A discussion for the case $N_\text{b}\eqt2$ is given in textbooks \cite{Kira2011}. 

\subsection{Gauge perspective of the {\eom}}
So far, we have derived equations of motion for density matrices, with examples focusing on homogeneous electric fields treated in Coulomb-gauge 
with $\bE = -\dot\bA$. 
Then, the operator relation \eqref{e30a} takes the form 
\begin{align}
    \ci\partial_t\rhoC(\bk;t) &= [\inh(\bk {-}\bA(t)),\rhoC(\bk;t)] 
    \label{e71} \,. 
\end{align}
In this section, we translate upper commutator relation 
into an {\eom} for matrix elements of $\rho(\bk;t)$. We represent $\rho(\bk;t)$ in two different basis sets and present the {\eom} associated with either one.

Within the co-moving basis~$|n\bk_t\brangle$, we have 
matrix elements $\varrho_{nn'}(\bk;t)$ 
\begin{align}
    \varrho_{nn'}(\bk;t) \coloneqq \blangle n\bk_t|\rho(\bk;t)|n'\bk_t\brangle\,,\label{e40a}
\end{align}
see Appendix~\ref{appha} where we show that  Eqs.~\eqref{e71} and~\eqref{e40a} indeed lead to the EoM~\eqref{e33a}.  

For exploring another basis, we define a \textit{boost operator} as 
\begin{align}
    \mfB(t) \coloneqq \mfBdef \label{e41c}
\end{align}
where the operator $\mfT$ keeps track of the proper ordering along the $\bk$-space trajectory; by definition, it acts on stationary Bloch states as
\begin{align}
    \mfB(t)\ \ketnk = \ketnksubt\,.\label{e43a}
\end{align}
For the case of a homogeneous electric field, we have $\dot\bk(t){=}\partial_t (\bk{-}\bA(t)){=}{-}\dot\bA(t)$ such that functions are shifted as $\mfB(t)f(\bk){=}f(\bk{-}\bA(t))$ (see Appendix~\ref{appga}) in line with Eq.~\eqref{e43a}.
One may interpret the boost operator as analogon to the generator of translation that is a function of the momentum operator.
 By applying the boost operator
 \begin{align}
     \inh(\bk-\bA(t)) = \mfB(t) h^\text{in}(\bk) \mfB^{-1}(t) 
     \label{e43b}
 \end{align}
we translate the initial, unperturbed Hamiltonian~$\inh(\bk)$ to the time-dependent Hamiltonian~$h(\bk;t)$ at time~$t$.
 
 With the definition of the density matrix in the \textit{dipole gauge}~\footnote{
We have discussed the Coulomb gauge (that is also referred to as velocity gauge) and dipole gauge (that is also referred to as length gauge) in footnote~\onlinecite{Note1}. 
The eigenstates of a Hamiltonian in both gauges are connected by a space-time dependent  transformation~\cite{Gottfried1966, Landau1981} as it is also used in recent work~\cite{Foeldi2017} focusing on dynamics of Bloch electrons.
The space-dependence of this transformation turns into a derivative in $\bk$ such that eigenstates of a Hamiltonian in both gauges transform via the Boost operator~$\mfB(t)$.
As consequence, the density matrix in both gauges transforms as in Eq.~\eqref{e73}.}
\begin{align}
     \rhoD(\bk;t)
\coloneqq\mfB^{-1}(t) \rhoC(\bk;t)\mfB(t)
     \label{e73} 
 \end{align}
 one can derive an EoM from Eq.~\eqref{e71} as
 \begin{align}
     \ci[\partial_t + \dot\bk(t) \partial_\bk]\  \rhoD(\bk;t) = [h^\text{in}(\bk),\rhoD(\bk;t)]\,.
     \label{e74} 
 \end{align}
In this representation, the commutator involves the unperturbed Hamiltonian only. It therefore is evaluated conveniently in the stationary basis $|n\bk\brangle$. 
Similarly to Appendix~\ref{appha}, one derives the traditional dipole-gauge formulation of the SBE with the characteristic gradient term on the lhs,~\cite{Golde2008}
\begin{align}
\begin{split}
    &\ci [\partial_t+\cq\bE(t)\partial_\bk]\rhoDnnprime(\bk;t) =  \epsilon_{nn'}(\bk)\rhoDnnprime(\bk;t) 
    \\[0.3em]
    & 
    + \bE(t)\sum_{\un}
    \big(\rhoDnun(\bk;t) \mathbf{d}_{\un n'}(\bk)
    -  \mathbf{d}_{n\un}(\bk)\rhoDunnprime(\bk;t)
    \big)\,,
    \end{split}
    \label{e45}
\end{align}
using the definition
\begin{align}
\rhoDnnprime(\bk;t) &\coloneqq \brank \rhoD(\bk;t)\ketnk\,.
\end{align}
As shown in Appendix~\ref{appga}, this definition relates to the Coulomb-gauge density matrix elements in the co-moving basis from Eq.~\eqref{e40a} via
\begin{align}
    \varrho_{nn'}(\bk;t) = \rhoDnnprime(\bk{-}\bA(t);t)\,.\label{e48d}
\end{align}

An alternative way to derive Eq.~\eqref{e45}
starts from the dipole (or length) gauge 
in which the electric field is represented by a linear potential. The relation Eq.~\eqref{e73} between $\rhoC(\bk)$
and $\rhoD(\bk)$ is thus understood as a gauge transformation. 
We emphasize that the time evolution of physical observables resulting from the {\sbe} is gauge-independent,
of course.~\cite{Foeldi2017, Li2019}

\subsection{Phenomenological dephasing}

The formalism developed thus far has 
neglected the collision term 
$ \partial\rho/\partial t|_\text{coll}\eqt0$ in Eq.~\eqref{e3}; its most important physical effect is to provide a dephasing mechanism. The strength of the {\sbe} is that dephasing can be included  phenomenologically in Eq.~\eqref{e33a} by adding 
 a term that is damping oscillations of offdiagonal density matrix elements~\cite{Schaefer2002,Floss2018}: 
\begin{align} 
\begin{split}
\Big( &\ci \frac{\partial}{\partial t} +  \frac{\ci(1-\delta_{nn'})}{T_2}
-\epsilon_{nn'}(\bk_t)\Big)
\varrho_{nn'}(\bk;t) =\\[0.3em]
&\bE(t) 
\sum_{\un} 
\varrho_{n\un}(\bk;t)\bfd_{\un n'}(\bk_t)
-\,\bfd_{n\un}(\bk_t)\varrho_{\un n'}(\bk;t)\,.
\end{split}\label{e94}
\end{align} 
The damping translates to the \eom~\eqref{e45} in the stationary basis with the dipole gauge:
\begin{align}
    &\ci \left[\frac{\partial}{\partial t}+\frac{1{-}\delta_{nn'}}{T_2}+\cq\bE(t)\frac{\partial}{\partial\bk}\right]\rhoDnnprime(\bk;t) =  \epsilon_{nn'}(\bk)\rhoDnnprime(\bk;t) \nonumber
    \\[0.3em]
    & 
    + \bE(t)\sum_{\un}
    \big(\rhoDnun(\bk;t) \mathbf{d}_{\un n'}(\bk)
    -  \mathbf{d}_{n\un}(\bk)\rhoDunnprime(\bk;t)
    \big)\,.\label{e99}
\end{align}
While the damping term breaks the time reversal invariance, it respects particle number conservation and the gauge symmetries. 
In particular, 
$\sum_{n}\rho_{nn}(\bk;t)$ continues to be stationary.

We here follow previous authors\cite{Floss2018,Hohenleutner2015,Luu2015,Vampa2014,Yu2016,Baykusheva2020,Kilen2020} and consider the relaxation time approximation as a convenient and computationally efficient approach to mimic qualitatively many-body effects leading to dephasing. The approximation associates the same rate parameter with all components of the density operator however, with consequences for quantitative estimates that are hard to predict. While it is common practice to use the rate as a fitting parameter so as to diminish quantitative discrepancies with reference data, the overall procedure is to be taken with a grain of salt.

\section{Observables: Emission intensity, dynamical polarization and current} \label{sec:observables}

As a response to the time-dependent perturbing fields, the charge density is accelerated; it varies in time and therefore irradiates light. 
The calculation of the emitted light intensity starts from the familiar equivalence between longitudinal current density and the derivative of the polarization,~\cite{Griffiths1999,Jackson2009,Schaefer2002} 
\bea
 \bJ(t) =  \partial_t {\bf P}(t)\,. \label{e37} 
\eea
Experiments measure the  frequency resolved emission intensity~$I$, which is given by~\cite{Jackson2009}
\begin{align}
I(\omega) &=
\frac{\omega^2}{3c^3}\,|\bJ(\omega)|^2\,. \label{e39b}
\end{align}

In the following, we derive expressions for the (dynamical) polarization and the current of the emitted radiation.

\subsection{Dynamical polarization~$\bP$}
We compute the polarization~\footnote{In the case of a slowly variating 
electric field over the unit cell, the electric field is approximated to be constant and the perturbation is connected  to
the polarization: $\langle \hat{V}\rangle = \bE \cdot \mathbf{P}(t)$. The expression for the
perturbation is expanded  in the first order of $\bq$ to derive the expectation value of the dipole-operator. The zero 
order can be gauged out and is
neglected.}  as expectation value of the dipole operator $\cq\br$ in a general basis~$|\alpha\rangle$ from \eqref{e1} as~\cite{Schaefer2002}
\bea
{\bf P}(t) &=& \frac{1}{\cV} \text{Tr} \left[\cq \br\rho(t)\right]  = \frac{1}{\cV}\sum_{\alpha,\beta} \langle \alpha | \cq\br|\beta \rangle \rho_{\beta\alpha}(t) 
\nonumber\\
&=&\frac{1}{\cV}\sum_{mm'}\sum_{\bk\bk'} \langle m\bk|  \cq\br|m'\bk' \rangle \rho_{m'm}(\bk'\bk;t)\,,
\label{e39a}
\eea
with 
$\cV$ denoting the normalization volume.
Adopting the notation from Eq.~\eqref{e15}, we employ a basis~$|m\bk\rangle$ with $\bk$-independent lattice-periodic part $|m\brangle$,
  \begin{align}
      \langle \br|m\bk\rangle = \frac{1}{\sqrt{\cN}}\, e^{\ci\bk\br} \blangle \br|m\brangle \,. \label{e75b}
  \end{align}
 The major advantage of the~$|m\brangle$-basis over a $\bk$-dependent lattice-periodic part~$|n\bk\brangle$ is that gradient-terms in $\bk$ can be much easier handled. 
We also derive our main result for~$\dtP$ using a $\bk$-dependent lattice-periodic~$|n\bk\brangle$ basis in Appendix~\ref{aH}.
We keep the full $\bk$-dependence of
\[
\rho_{mm'}(\bk\bk';t) = \langle m\bk|\rho(t)|m'\bk'\rangle
\]
in \eqref{e39a} to properly account for $\bk$-derivatives later on.
We evaluate the dipole matrix element $\langle m\bk| \br|m'\bk' \rangle $ appearing in the polarization \eqref{e39a} adopting \eqref{e:app418} as
\begin{align} 
	\langle m\bk| \br |m'\bk' \rangle 
	=\frac{(2\pi)^d}{\cV} 
	\blangle m |\left[\ci\partial_\bk e^{-\ci( \bk-\bk')\br}\delta(\bk{-}\bk')\right]|m' \brangle\,.\label{e41a}
\end{align} 
With \eqref{e41a} and results from Appendix~\ref{a0}, we obtain
{\allowdisplaybreaks
\begin{align} 
 \bP(t) 
&= \cq \sum_{mm'}\intbzdk\intbzdkpiprime\ \rho_{m'm}(\bk'\bk;t) \nonumber
\\
&\hspace{3em} 	\times\blangle m|\left[\ci\partial_\bk e^{-\ci( \bk-\bk')\br}\delta(\bk{-}\bk')\right]|m'\brangle\,.\nonumber\\
&= \ci \cq  \sum_{mm'} 
\intbz \frac{d\bk}{(2\pi)^d}\ 
 \blangle m|m'\brangle\left.\frac{\partial\rho_{m'm}(\bk^\prime\bk;t)}{\partial 
\bk^\prime}\right|_{\bk'\to\bk}\nonumber\\
&= \ci \cq  \sum_{m} 
\intbz \frac{d\bk}{(2\pi)^d}\ 
  \left.\frac{\partial\rho_{mm}(\bk^\prime\bk;t)}{\partial 
\bk^\prime}\right|_{\bk'\to\bk}\nonumber\\
&= \ci \cq   
\intbz \frac{d\bk}{(2\pi)^d}\ 
\text{Tr}_n\, \brho(\bk;t)\,,
\label{e42}
\end{align}}where integration by parts has been used to arrive at the second equation. 
In the last line, we defined 
\be 
\brho_{mm'}(\bk;t) \coloneqq \langle c^\dagger_{m'\bk}\partial_{\bk} c_{m\bk}\rangle =
\left.\frac{\partial\rho_{mm'}(\bk^\prime\bk;t)}{\partial 
\bk^\prime}\right|_{\bk'\to\bk}\label{e41}\,.
\ee 
The trace in Eq.~\eqref{e42} can be evaluated in any lattice periodic basis and it is our preferred choice to continue with basis-independent representations.
For computing the emission from Eq.~\eqref{e39b}, we employ the time derivative of~$\bP$ that translates to the time derivative of~$\brho$ in Eq.~\eqref{e42}.
We insert EoM  \eqref{e71}, $\ci \dot \rho(\bk;t)\eqt [h(\bk;t), \rho(\bk;t)] $ in the Coulomb gauge  in the rhs of~\eqref{e41} and obtain
\begin{align}
\ci \text{Tr}_n \dot\brho(\bk;t) = \text{Tr}_n \Big[ [h(\bk;t),\brho(\bk;t)] + (\partial_\bk h(\bk;t)) \rho(\bk;t)  
\Big]. \end{align} 
Since the trace of the commutator vanishes, we have 
 \begin{align}
     \dtP(t) &= \cq   
\intbz \frac{d\bk}{(2\pi)^d}\ 
\text{Tr}_n \Big[(\partial_\bk h(\bk;t))\, \rho(\bk;t) \Big]\,.
\label{e42c}
 \end{align}
We evaluate the trace in Eq.~\eqref{e42c} 
for the special case of a homogeneous electric field,  \mbox{$h(\bk;t)=\inh(\bk{-}\bA(t))$}
in the co-moving basis \mbox{$\ketnkt=\ketnksubt$}: 
\begin{align} 
\hspace{-1em} \dtP(t) 
&=
\cq  \sum_{n n'}  \intbzdkpi\ 
\blangle n\bk;t|
\frac{\partial \inh(\bk{-}\bA(t))}{\partial\bk}
|n'\bk;t\brangle\ \varrho_{n'n}(\bk;t)
\label{e46}
\\
&=
\cq  \sum_{n n'}  \intdkpi\ 
\blangle n\bk_t|
\frac{\partial \inh(\bk_t)}{\partial\bk_t}
|n'\bk_t\brangle\ \varrho_{n'n}(\bk;t)
\label{e95}
\end{align}
so that the density matrix~$\varrho_{nn'}(\bk;t)$ as defined in Eq.~\eqref{e40a} in the co-moving basis appears. In this way, it is possible to use~$\varrho_{nn'}(\bk;t)$ from the dynamics in Eq.~\eqref{e33a} to evaluate $\dtP(t)$ and subsequently also the emission intensity. 

The transparent result 
\eqref{e46} implies that the velocity associated with the co-moving states $|n\bk;t\brangle$ as given by the matrix element derives from the {\it instantaneous} band structure. 
Notice, however, that this particular aspect of 
\eqref{e46} is a consequence of our choice of gauge. In the later Section 
\ref{ss.IIIC} an equivalent expression, Eq.~\eqref{e85}, will be derived for the current density that involves the unperturbed band-structure.

\subsection{Longitudinal current density $\bJ$}

An alternative derivation of \eqref{e42c} embarks on the relation~\eqref{e37} between the longitudinal charge current density and the polarization, $\bJ(t)= \dot \bP$ and
\begin{align}
\bJ(t) &= \frac{1}{\cV} \text{Tr} \left[\cq \dot\br\rho(t)\right]=\frac{1}{\cV}\sum_{\alpha,\beta} \langle \alpha | \cq \dot \br |\beta \rangle \rho_{\beta\alpha}(t) 
\nonumber\\
&=\cV \sum_{nn'} \iintbzdkpi\  
\langle n\bk|\cq \dot \br | n'\bk'\rangle \rho_{n'n}(\bk'\bk;t)\,.\label{e60a}
\end{align}
Since the velocity operator $\dot \br$ relates to the Hamiltonian via the operator derivative 
 $\dot \br = \partial h/\partial \bp$, we readily conclude 
\be 
\bJ(t) = \cq \cV  \sum_{nn'} \iintbzdkpi\ 
\langle n\bk|\frac{\partial h(t)}{\partial \bp}| n'\bk'\rangle \rho_{n'n}(\bk'\bk,t) \,.
\label{e511} 
\ee 
\subparagraph{Translational invariance:} In the special situation of translational invariance, $h$ is diagonal in the eigenstates $|n\bk\rangle$ of the momentum operator $\bp$. Therefore, first the operator derivative $\partial/\partial \bp$ in Eq. \eqref{e511} can be replaced by $\partial/\partial\bk$ and second, the matrix element is proportional to $\delta(\bk-\bk')$.~\cite{Xiao2010} Hence,
Eq. \eqref{e511} simplifies 
to 
\begin{align}
\bJ(t) &= \cq  
\intbzdkpi \ 
\text{Tr}_n \left[
\frac{\partial h(\bk;t)}{\partial \bk}
 \,\rho(\bk;t)\right]\,.
\label{e60}
\end{align}
and we recover \eqref{e42c}. 
%

\subsection{\label{ss.IIIC} Inter- and intraband currents:  Anomalous velocity, conductivity tensor and damping current}
For additional physical insight, we  split the current density $\bj(t)$ into semiclassical and quantum contributions.
To this end, we embark on the trace formula Eq.~\eqref{e42c}
\begin{align}
\bj(t)
=  \cq   
\intbz \frac{d\bk}{(2\pi)^d}\ 
\text{Tr}_n \Big[(\partial_\bk h(\bk;t))\, \rho(\bk;t) \Big]\,.\nonumber 
\end{align} 
Recalling Eq.~\eqref{e73} and \eqref{e43b}, 
we derive an expression in the dipole gauge as
\begin{align}
\bj(t)
&= \cq \intbzdkpi 
\  \text{Tr}_{n}\Big[ \mfB^{-1} \,\partial_\bk  h^\text{in}(\bk{-}\bA(t))\,\mfB\, \rhoD(\bk;t)\Big]\nonumber\\
&= \cq \intbzdkpi  
\  \text{Tr}_{n} \Big[\mfB^{-1}\, \partial_\bk  \mfB\, h^\text{in}(\bk)\, \rhoD(\bk;t)\Big]\nonumber\\
&= \cq \intbzdkpi  
\  \text{Tr}_{n}  \Big[\partial_\bk h^\text{in}(\bk)\, \rhoD(\bk;t)\Big]\label{e84}
\end{align} 
where the last line is assuming $\dot \bk$ does not depend on $\bk$, as is the case for homogeneous electric fields. The trace in Eq.~\eqref{e84} when evaluated in the stationary basis~$\ketnk$ yields a formula 
\begin{align}
 \bj(t)  =
\cq  
\sum_{nn'} \intbzdkpi
\ \brank  \partial_\bk h^\text{in}(\bk)\ketnkprime\ \rhoDnprimen(\bk;t)\,,\label{e85}
\end{align} 
which has frequently been used before\cite{Aversa1995,Sipe2000,AlNaib2015,McGouran2016,McGouran2017, Floss2018,Li2019,Noda2019,Chan2019,Yue2020,Yue2020b,deJuan2020,Chacon2020}.

In Appendix~\ref{apph} we derive an expression for the matrix element 
\begin{align}
   \brank \partial_\bk h^\text{in}(\bk)\ketnkprime = \delta_{nn'}\partial_\bk\epsilon_n(\bk) 
   -\frac{\ci}{\cq} \mathbf{d}_{nn'}(\bk)\epsilon_{n'n}(\bk)\label{e86}\,.
\end{align}
Inserting~\eqref{e86} into \eqref{e85}, we can motivate the  splitting of \eqref{e85} into intraband ($n{=}n'$) contributions and a rest  ($n{\neq}n'$). 
We  reproduce a frequently used expression for the intraband current,~\cite{\refsemissionformula}
 \begin{align}
     \jintratilde 
   & \coloneqq  \cq
  \sum_{n} \intbzdkpi\ \partial_{\bk}\epsilon_n(\bk)\
  \rhoDnn(\bk;t)\,,\label{e87}
 \end{align}
that adds together with the interband current 
\begin{align}
 \jintertilde & \coloneqq
\cq  
\sum_{n\neq n'} \intbzdkpi
\ \brank  \partial_\bk h^\text{in}(\bk)\ketnkprime\ \rhoDnprimen(\bk;t) \label{e69a}
\\ 
&\hspace{-1em}=-\ci 
  \sum_{n\neq n'} \intbzdkpi\
  \mathbf{d}_{nn'}(\bk)\ \epsilon_{n'n}(\bk)
  \rhoDnprimen(\bk;t)\label{e75}
\end{align} 
to the total current
   \begin{align}
   \bj(t) = \jintratilde + \jintertilde\,.
   \label{e70a}
  \end{align}
  Embarking on \eqref{e45}, 
 we can also write 
  \begin{align} 
  &\jintertilde = 
  \sum_{n\neq n'} \intbzdkpi\
  \mathbf{d}_{nn'}(\bk)\
  (\partial_t+ \cq \bE(t)\partial_\bk)
  \rhoDnprimen(\bk;t)\,
  \nonumber\\
  & -\ci\sum_{n\neq n'} \intbzdkpi\
  \mathbf{d}_{nn'}(\bk)\
  [\bE(t)\mathbf{d}(\bk),\rhoD(\bk;t)]_{n'n}\,.
    \label{e68} 
 \end{align}
The first term in \eqref{e68} has the interpretation of a polarization current\cite{\refsemissionformula,Aversa1995}, 
  \begin{align}
   \dtPinter\coloneqq
  \sum_{n\neq n'} \intbzdkpi\ {\bf d}_{nn'}(\bk)\ \rhoDdotnprimen(\bk;t)\,.
  \label{e69} 
 \end{align}
As we show in  Appendix~\ref{sec:appg}, the remaining two terms in \eqref{e68} have a natural splitting into two parts:
The first part adds to the intraband current  $\jintratilde$ and accounts for the anomalous contribution to the (semiclassical) velocity\cite{Xiao2010}: 
\begin{align}
    \mathbf{v}_n(\bk) = \partial_\bk\epsilon_n(\bk) + \cq\bE(t)\times\bomega_n(\bk)\,;
\end{align}
 the Berry curvature~$\bomega_n(\bk)$ is given in three dimensions as~\cite{Xiao2010}
\begin{align}
\bomega_n(\bk) &=  \frac{1}{\cq}  \boldsymbol\nabla_\bk\times \mathcal{A}_n(\bk)\,,
\label{e72a}
\end{align}
with $\mathcal{A}_n(\bk)\eqt\bfd_{nn}(\bk)$ as defined in Eq.~\eqref{e33b}.
So, the full intraband current reads 
\begin{align}
\jintra = 
 \cq
  \sum_{n} \intbzdkpi\ \mathbf{v}_n(\bk)\
  \rhoDnn(\bk;t)\,.\label{e72}
  \end{align}
The second part takes the form $\sigma(t) {\bf E}(t)$ with 
\begin{align}
    \sigmamatel =  \sum_{n\neq n'} \intbzdkpi&
    \bigg[
    \ci\,d_{nn'}^{(j)}(\bk)
    \left(d_{nn}^{(i)}(\bk)-d_{n'n'}^{(i)}(\bk)\right)
    \nonumber
    \\
    & \hspace{-1em}
    -\left(\cq\partial_{k_i}d_{nn'}^{(j)}(\bk)\right)
    \bigg]
    \rhoD_{n'n}(\bk;t)\,.
    \label{e70}
\end{align}
Gauge invariance with respect to multiplicative wavefunction phase factors~\cite{Li2019} can be easily shown for the conductivity tensor~\eqref{e70}.
Collecting terms, we  have for the current 
 \begin{align}
     \bJ(t) = \jintra + \dtPinter + 
     \sigmainter \bE(t)  \,.
     \label{e59c}
\end{align}
When deriving Eq. \eqref{e68} and therefore \eqref{e59c} we have employed the \eom~\eqref{e45}, i.e.~we have not accounted for phenomenological damping terms. The latter can be included by using \eom~\eqref{e99} instead of \eqref{e45}. The effect of dephasing amounts to an effective contribution   
\begin{align}
    \Jdamp \coloneqq \frac{1}{T_2}\sum_{n\neq n'}\intbzdkpi\, \bfd_{nn'}(\bk)\,\rhoD_{n'n}(\bk;t)
    \label{e95a}
\end{align}
that relates to the polarization current~\eqref{e69} via
\begin{align}
    \dtPinter = T_2\,\partial_t\,\Jdamp
    \label{e81}
\end{align} and adds to the previous result 
\eqref{e59c}. 
Summarizing, we have for the total current a splitting in intraband and interband contributions
 \begin{align}
     \bJ(t) = \jintra + \jinter
     \label{e59}
\end{align}
defining the interband current as
 \begin{align}
     \jinter  \coloneqq \dtPinter + 
     \sigmainter \bE(t) + \Jdamp \,.
     \label{e96}
\end{align}
Further details of the derivation of  Eq.~\eqref{e59}/\eqref{e96} are given in Appendix~\ref{sec:appg}.

\subsection{Work deposited by $\bE(t)$: the case of two-bands }
As a first application of our result  \eqref{e59}, we derive an expression for the electric work deposited per time $\dot W(t)\,{\coloneqq}\,\bj(t)\bE(t)$.
We begin with the observation that the anomalous contribution to the charge current is of the form 
$\bE\,{\times}\,\bomega_n$ and therefore does not contribute to $\dot W(t)$. We therefore can adopt \eqref{e87} and \eqref{e75} for intra- and interband current contributions. 

We focus on a two-band model for a band insulator. With respect to dynamics we thus deal with a two-level model with a conservation law: $\rhoD_{vv}(\bk;t){+}\rhoD_{cc}(\bk;t)\eqt z(\bk)$ for valence ($v$) and conduction ($c$) band. The $x$-axis is taken to point along the electric field; we then have for the intraband current
\begin{align}
    \jintrax =
    \cq\intbzdkpi\, 
    \frac{\partial\epsilon_{cv}(\bk)}{\partial k_x}\,
    \rhoD_{cc}(\bk;t)\label{e89}\, + j^\text{eq}_x, 
\end{align}
with the equilibrium current
$$
j^\text{eq}_x\coloneqq- \int \frac{d\bk}{(2\pi)^d} \, \epsilon_{v}(\bk)\, \frac{\partial z(\bk)}{\partial k_x}\,; 
$$the interband current reads
\begin{align}
\jinterx   &=
2 \intbzdkpi\, \epsilon_{cv}(\bk)\,
\text{Im} \Big(d_{vc}^x(\bk)\,\rhoD_{cv}(\bk;t)\Big)\,.
\end{align}
%
Recalling the {\eom}~\eqref{e99} and using integration by parts, we decompose the interband current into
\begin{align}
\jinterx &=
 \intbzdkpi\, \epsilon_{cv}(\bk)
\left[
\cq \frac{\partial}{\partial k_x}
+
\frac{1}{E(t)}\frac{\partial}{\partial t}
\right]
\rhoD_{cc}(\bk;t)
\\
&=j_x^\text{eq}-\jintrax + 
\intbzdkpi\,
 \,
\frac{\epsilon_{cv}(\bk)}{E(t)}\,
\frac{\partial\rhoD_{cc}(\bk;t)}{\partial t}\,.
\end{align}
For the total current~$j_x(t)\eqt\jintrax{+}\jinterx$, we thus arrive at  
\begin{align}
       j_x(t)-j_x^\text{eq} = \frac{1}{E(t)} 
\intbzdkpi\,
\epsilon_{cv}(\bk) \,
\frac{\partial\rhoD_{cc}(\bk;t)}{\partial t}\,.
\label{e88}
\end{align}
This expression implies 
\begin{align}
    \dot W(t) =  
\intbzdkpi\,
\epsilon_{cv}(\bk) \,
\partial_t\left[\rhoD_{cc}(\bk;t)-\rhoD_{vv}(\bk;t)\right]/\,2\,,\label{e93a}
\end{align}
where $\dot W(t)\,{=}\,(j_x(t)-j_x^\text{eq})E(t)$ has been employed. Equation \eqref{e93a} represents a transparent result for the electric work done on the system per time: whenever a particle-hole pair is created at wavenumber $\bk$, an amount of energy $\epsilon_{cv}(\bk)$ is deposited into the system. Within this simple model alternative routes for energy deposition do not exist. 

\subsection{Relation to earlier work}
Sipe and coworkers\cite{Aversa1995,Sipe2000} have considered the current density $\bj(t)$ in their work on second and third order responses. In this context they arrived at a splitting of the total current density similar to \eqref{e59c}. 
Their derivation employs a perspective focusing on the polarization as central concept, especially in Ref.~\onlinecite{Sipe2000}.  
Correspondingly, it starts with a decomposition of the position operator into an inter- and intraband constituent. Our derivation is somewhat simpler, in the sense that no such decomposition is imposed at any time. The constituents of our final 
result \eqref{e59c} and their physical nature more or less reveal themselves in the course of our calculation. 

Frequently cited works\cite{\refsemissionformula} on high-harmonics generation have used approximate variants of~\eqref{e59}: 
 the anomalous velocity, $\sigmainter\bE(t)$, and $\Jdamp$  have not been accounted for.
It is important to note that
the anomalous term in the velocity as well as the $\sigma$-term both equal zero when  the following two conditions on $\bfd_{nn'}(\bk)$ are satisfied: all diagonal entries  vanish,~$\mathbf{d}_{nn}(\bk)\eqt0$ and the off-diagonals $\bfd_{nn'}$ are independent of $\bk$. 
Indeed, models for the dipole-matrix have frequently been adopted that satisfy these conditions~\cite{\refsemissionformula}; 
the main approximation for the current calculation in these works therefore is the neglect of $\Jdamp$. 
%


\section{Application: Dynamics of Dirac fermions} \label{sec5}
Motivated by recent experiments~\cite{Yoshikawa2017,Hafez2018,Higuchi2017,Heide2018,Heide2019,McIver2020,Cheng2020,Kovalev2020,Lim2020},  we briefly present an application of the {\sbe} formalism to the density matrix dynamics for a Dirac-type dispersion driven by  an ultra-short electric field pulse. We focus on bandstructure effects and  neglect mean-field interactions.

\subsection{Model and method}

\subparagraph*{Hamiltonian.}
We employ a two-dimensional Dirac cone
\begin{align}
    \inh(\bk) = \cone \label{e84a}
\end{align}
with a Fermi velocity $\vF\eqt4.3\cdot 10^5\,\text{m/s}\eqt1.44\cdot10^{-3}c$ that is a prototypical two-band surface Hamiltonian of a topological insulator as bismuth telluride (Bi$_2$Te$_3$).~\cite{Liu2010}
Such a model Hamiltonian can be obtained, e.g., from \textit{ab-initio} calculations by~$\bk\cdot\mathbf{p}$ perturbation theory~\cite{Liu2010} or the use of Wannier functions~\cite{Osika2017,Silva2019b}.
The eigenstates and bandstructure are computed as
\begin{align}
\begin{split}
&| v\bk\brangle = \frac{1}{\sqrt{2}}\left(
\begin{array}{c}
      1  \\[0.2em]
     \ci e^{\ci\theta}
\end{array}
\right)\,, \hspace{1.5em}
| c\bk\brangle = \frac{1}{\sqrt{2}}\left(
\begin{array}{c}
      -1  \\[0.2em]
     \ci e^{\ci\theta}
\end{array}
\right)\,, 
\\[0.8em]
&\epsilon_v(\bk)=-\vF|\bk|\,,
\hspace{3.5em}
\epsilon_c(\bk)= \vF|\bk|\,,
\end{split}
\label{e91a}
\end{align}
for~$v$ and $c$ being the valence and conduction band, respectively.
The dipoles follow
\begin{align}
  &\mathbf{d}_{nn'}(\bk) = -  \frac{\cq}{2|\bk|}  
  \,\hat{e}_\theta
 \label{e93} 
 \end{align}
for~$n,n'\,{\in}\,\{v,c\}$  with $\theta$ being the polar angle and $\hat{e}_\theta$ the unit vector  orthogonal to~$\bk$.

\subparagraph*{Electric-field pulse.}
An ultra-short laser pulse is employed with an electric driving field that is polarized in $x$-direction,
\begin{align}
    \bE(t)  = E\,\hat{e}_x \sin(\omega_0 t)\, \exp\left(-\frac{t^2}{\sigma^2}\right)\,,\label{e91}
\end{align}
where $\omega_0\eqt2\pi\,{\cdot}\,25\,\text{THz}$, $E \eqt 5\,\text{MV/cm}$ and~$\sigma\eqt50$\,fs through\-out our calculations. 
The pulse shape here adopted follows the experimental ones.~\cite{Schubert2014,Hohenleutner2015,Langer2016}

\subparagraph*{Equations of motion.} 
The {\eom} will be adopted from Eq.~\eqref{e94} (Coulomb gauge) and~\eqref{e99} (dipole gauge).
For practical calculations, we have chosen~$T_2\eqt1\,\text{fs}$ following Ref.~\onlinecite{Hohenleutner2015}, similar to Refs.~\onlinecite{Luu2015,Vampa2014,Yu2016,Baykusheva2020}; for further discussion see Ref.~\onlinecite{Floss2018}.
The initial condition for integrating the {\eom} was chosen with the valence band being filled  and  the conduction band being empty: 
\begin{align}
    \inrho_{nn'}(\bk) = \delta_{n\text{v}}\delta_{n'\text{v}}. 
\end{align}
For the $\bk$-domain of integration, we have allowed for the limit $\pi/a\,{\to}\,\infty$. 
This corresponds to a situation where the linear 
dimension of the simulation volume, $k_\text{max}{\coloneqq}\pi/a$, exceeds any other characteristic wavenumber; in particular,  $\pi/a\,{\gg}\,\omega_0/\vF$ and $\pi/a\,{\gg}\, \cq E/\omega_0$. 
The inequalities are satisfied for typical lattice constants $a$ and parameters~$\omega_0,\vF, E$ as chosen in this work:
For $a\eqt 3\,$\AA, we have  $\pi/a\,{\simeq}\, 2\, \cq E/\omega_0$ such that Bloch electrons excited at the $\Gamma$-point hardly reach the boundary $k_\text{max}$ of the simulation volume.
Much higher field strengths up to 72\,MV/cm are used to drive the Bloch electrons beyond $k_\text{max}$ to initiate Bloch oscillations~\cite{Schubert2014}.
For integrating the {\eom} in Eq.~\eqref{e94}, we use a backward differentiation formula with a maximum adaptive timestep of 0.1\,fs as implemented in scipy~\cite{Virtanen2020}.
Convergence with respect to the $k$-point mesh size as well as numerical equivalence of  Coulomb and dipole gauge is demonstrated in Appendix~\ref{sec:apph}, 
especially Fig.~\ref{fig:DiracwithoutB}\,(d). 
The emission intensity $I(\omega)$ is computed from Eq.~\eqref{e39b}  using the current density from Eq.~\eqref{e95} (Coulomb gauge) or Eq.~\eqref{e85} (dipole gauge).
For the simulations, we have used our in-house program
package CUED, freely available from github, \url{https://github.com/ccmt-regensburg/CUED}.

\subsection{Results: Dynamics in homogeneous $\bE$-field}\label{subsec:5C}
 
\begin{figure}[]
    \centering
\includegraphics[width=8.5cm]{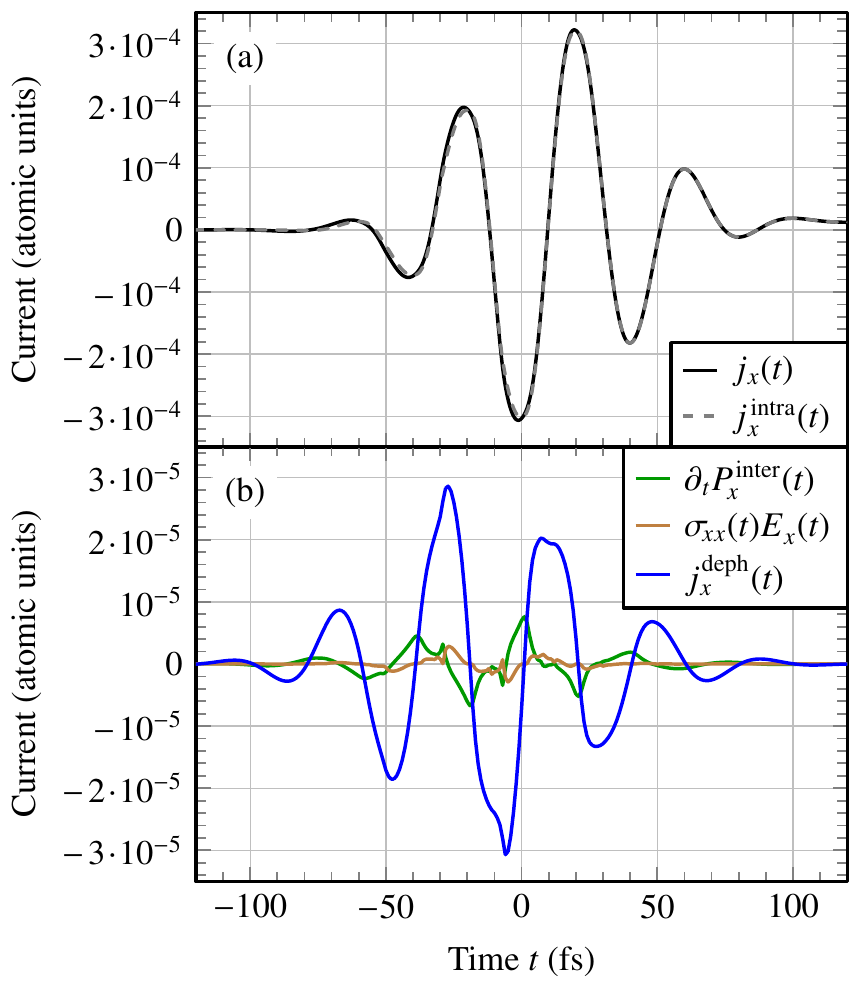}
    \caption{Time-dependent currents for the Dirac system. 
    (a)~$x$-component of the  total current (black, computed from Eq.~\eqref{e85}) and the  intraband current (dashed gray, computed from Eq.~\eqref{e87} using $\jintrax\eqt\jintratildex$ due to $\bE(t)\eqt E_x(t)\hat{e}_x$).
(b) $x$-component of interband currents, 
$\dtPinter$ is computed from~\eqref{e69}, $\sigma(t)\bE(t)$ from \eqref{e70}, $\Jdamp$ from \eqref{e95a}  and $\jintertilde$ from \eqref{e69a}.}
    \label{fig:current}
\end{figure}

{\bf Real-time currents.} In Fig.~\ref{fig:current}\,(a) we display the in-plane current component, $j_x(t)$, directed along the electric field. For the parameter regime here considered, the current in Fig.~\ref{fig:current} is dominated by the semiclassical contribution $\jintrax$ with a shape that roughly follows the vector potential. 
The deviations of $\jintrax$ from the full current are seen to be largest at early times. The reason is that the system we consider starts out with the valence band being fully occupied and the conduction band being empty; the semiclassical current can start to flow only after occupations of conduction band electrons (and valence band holes) have built up.

%

%
%
%
%

\begin{figure}[]
    \centering
\includegraphics[width=8.5cm]{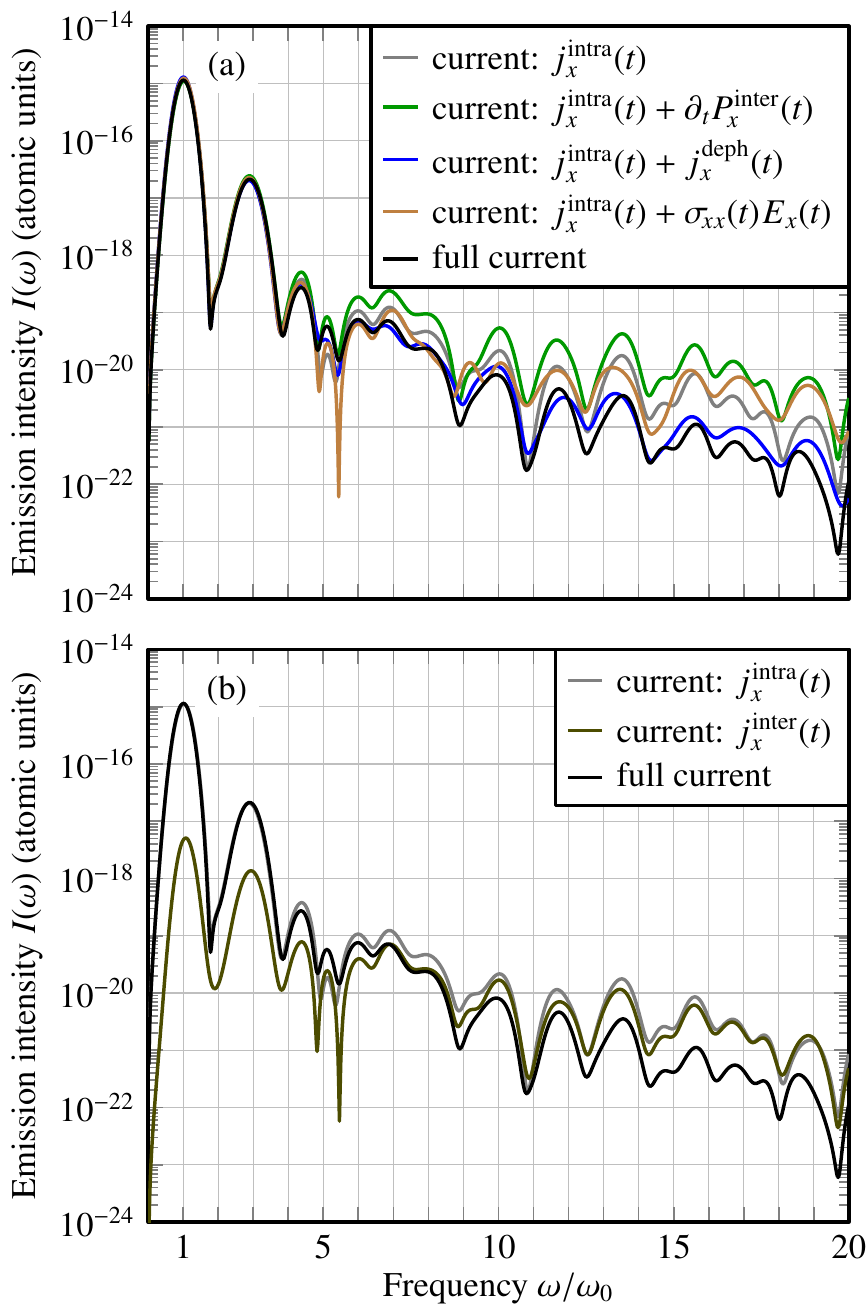}
    \caption{ Composition of the emission intensity~$I(\omega)$, Eq.~\eqref{e39b}, for various combinations of currents; the plot illustrates the relative importance of different contributions.  
    (a)
    Gray: intraband current~$\jintrax$ from Eq.~\eqref{e72} only,
    blue: sum of~$\jintrax$  and polarization related current~$\dtPinterx$ from Eq.~\eqref{e69},
    green: sum of~$\jintrax$ and current~$\jdampx$ due to dephasing from Eq.~\eqref{e95a}, 
    blue: sum of~$\jintrax$ and current~$\sigma_{xx}(t)E(t)$ due to the conductivity tensor from Eq.~\eqref{e70}, 
    black: full current using Eq.~\eqref{e85}.
   (b)
   Comparison of intraband and (full) interband current, $\jinterx\eqt\dtPinterx\pt\jdampx\pt\sigma_{xx}(t)E(t)$, c.f. Eq.~\eqref{e69a}.
    }
    \label{fig:emission}
\end{figure}

The discrepancy between the total current and the intraband current is due to interband currents that are shown in Fig.~\ref{fig:current}\,(b); for a numerical check of the discrepancy see Appendix~\ref{sec:apph}.
For strong damping, $T_2\eqt1$\,fs, the dephasing contribution $\jdampx$
dominates the interband current, $\jinterx$, see Fig.~\ref{fig:current}\,(b), since
$\jdampx \propto T_2^{-1}$; 
in our case it exceeds the other terms, $\dtPinterx$ and $\sigma_{xx}(t)E_x(t)$, by nearly an order of magnitude.
In a sense, this observation also carries over to the high-harmonic generation: 
Fig.~\ref{fig:emission}\,(a)  shows that at high-harmonic order six and higher the emission falls below the value that it had were it only for the intraband current alone. Only upon adding the dephasing current, the emission decays by up to a factor of ten down to its real value.

We comment on the significance of this observation. Dephasing rates of order $T_2^{-1}\sim 10^{15}$Hz have frequently been employed in numerical investigations\cite{Hohenleutner2015,Luu2015,Vampa2014,Yu2016,Baykusheva2020,Floss2018}. One of the effects of strong dephasing is to dampen  oscillating terms in the {\sbe} and - consequently - also in physical observables, such as the current $\bj(t)$. As has been discussed by \textcite{Floss2018}, in a crude way this damping of fluctuations mimics the spatial averaging that occurs in experiments because different sample regions experience different strength of the laser field and therefore contribute incoherently to the experimental signal. 

Now as we have shown, adding phenomenological terms to the {\sbe}, in principle, gives an extra contribution to the charge current, $\jdampx$, that incorporates genuine many-body effects, such as friction. This term will not arise with spatial averaging; therefore, this term should be omitted in current calculations for the purpose of mimicry.  
Our results in Fig. \ref{fig:current} and \ref{fig:emission} emphasize the quantitative importance of this term at large damping and therefore 
underline a qualitative difference of spatial averaging from dephasing. 

The remaining contributions to the interband current, $\dtPinterx$ and $\sigma_{xx}(t)E_x(t)$,  fall below $\jintrax$ by two orders of magnitude. Since they are  considerably sharper structured than~$\jdampx$, see Fig.~\ref{fig:current}\,(b), they nevertheless contribute significantly to the high harmonics in Fig.~\ref{fig:emission}. 
We explain this finding with the fact that $\dtPinterx$ and $\sigma_{xx}(t)E_x(t)$ contain derivatives in time and $k$, respectively, in contrast to $\jdampx$.
In passing, we note that in Fig. \ref{fig:current} the extrema of~$\jdampx$ are seen to be the roots of~$\dtPinterx$. The correlation reflects the exact identity Eq. \eqref{e81}. 
%

{\bf Emission intensity.}
The emission spectrum is shown in Fig.~\ref{fig:emission}.
%
%
After an exponential decay by four orders of magnitude a nearly plateau-like regime is seen from the 5th to 20th harmonic order.
Similar behaviour has been reported in the literature for a  semimetallic Hamiltonian~\cite{Tamaya2016} and the Haldane model~\cite{Silva2019}. In  Appendix 
\ref{sec:appsemic}
we investigate a toy-model of a semiconductor, which also reproduces the plateau feature, see Fig.~\ref{fig:emissionsemic}.

%
%
In Fig.~\ref{fig:emission}\,(b) the intraband current dominates the emission spectrum up to the fifth harmonic order. This observation is in line with previous studies on semiconductors\cite{Vampa2014} and implies that the lower frequency response is essentially semi-classical also for the Dirac system. 
At higher harmonic orders~($\omega/\omega_0\gt10$) the intra- and interband currents develop a similar strength. Remarkably, they interfere destructively, so that the combined transmission is smaller by up to order of magnitude as compared to the individual ones. This special feature discriminates the Dirac-cone from the semi-conductor,  see Fig.~\ref{fig:emissionsemic} in Appendix~\ref{sec:appsemic} for an semiconducting paradigm.

The increased importance of interband currents for high harmonics manifests already in the time-dependent currents in Fig.~\ref{fig:current} (b):  the interband currents~$\dtPinterx$ and~$\sigma_{xx}(t)E_x(t)$ feature sharp kinks adding strong weight to high-frequency amplitudes. 
%

\section{Conclusions and Outlook} 
A derivation of the semiconductor Bloch equations (\sbe) for the time evolution of the density matrix has been presented emphasizing the close  relation to the  Berry  connection.
This particular approach has the appealing feature that it  lends itself to  a  semiclassical  perspective  on  the  SBE  allowing for a simplified treatment of magnetic-field effects by Lorentz forces that will be presented in a forthcoming publication. 

Also, expressions have been rederived connecting the density matrix to physical observables, specifically, to the current density. In addition to the traditional current, summing intraband and interband-polarization contributions~\cite{\refsemissionformula}, we have identified an extra term; it becomes sizable in situations where dipole-matrix elements depend strongly on the wavenumber. 
We have implemented an SBE solver and applied it to Dirac metals, motivated by the observation that dipoles are strongly $\bk$-dependent for Dirac fermions. We find that the extra term  gives a significant contribution to the total current, in particular, to the high-harmonic generation: the emission intensity can deviate by more than an order of magnitude upon neglecting the extra term. 

\begin{acknowledgments}
We thank Martin Axt,  Paulo E.~de Faria Junior, Rupert Huber, Vanessa Junk, Christoph Lange, Sivan Refaely-Abramson, Klaus Richter and Mathias Steinhuber  for helpful discussions.
We thank an anonymous referee for an important hint in the process of deriving Eq.~\eqref{e96}.
Support from the German Research Foundation (DFG) through the Collaborative
Research Center, Project ID 314695032 SFB 1277
(project A03) is gratefully acknowledged.
J.~Crewse acknowledges funding by the NSF (National Science Foundation), project ID DMR-1828489.

\end{acknowledgments}

\appendix

\section{Basics of lattice periodicity}\label{a0}
We define~$\cV$ as a volume containing several unit cells with lattice vectors $\bR$.
Then, an integration over~$\cV$ is given by integrating over individual cells,
\be
\int_\mathcal{V} d\br \ f(\br) = \sum_{\bR} \int_{\cC} d\br\ f(\br+\bR) \,.\label{eqa1}
\ee
$\cC$ denotes the integration over the (primitive) unit cell. 

For vectors $\bk{-}\bk'$ from the first Brillouin zone, we recall  
\be
\sum_\bR e^{\ci (\bk-\bk')\bR} = {\cN}\delta_{\bk\bk'} 
\simeq \frac{(2\pi)^d}{{\mathcal V}_\text{c}}\delta(\bk - \bk') \label{eqa2}
\ee
where $\cN{\coloneqq}\sum_{\bR}$ denotes the number of unit cells in $\mathcal V$
and ${\mathcal V}_\text{c}={\mathcal V}/{\cN}$ is the volume of a unit cell. 
The rhs of \eqref{eqa2} in the limit of large~$\cN$  implies
\begin{align}
    \sum_{\bk}f(\bk) \, \simeq \frac{{\mathcal V}}{(2\pi)^d} \intbz d\bk \ f(\bk)\,,\label{eqa3a}
\end{align}
where we integrate over the first Brillouin zone.

The eigenstates of the stationary, lattice-periodic Hamiltonian are Bloch-states $|n\bk\rangle$.
In the context of Eq.~\eqref{e15}, we have defined the lattice periodic wavefunction as~$u_{n\bk}$:
\begin{align}
 \langle \br|n\bk\rangle = \frac{1}{\sqrt{\cN}}\, e^{\ci\bk\br} \blangle \br|n\bk\brangle  =: \frac{1}{\sqrt{\cN}}\,e^{\ci\bk\br}\,u_{n\bk}(\br)\,.
 \label{eqa3}
\end{align}
The double angular brackets indicate that the normalization volume for $u_{n\bk}$ is the unit cell~$\cC$:
\begin{align}
    \blangle n\bk|n'\bk\brangle \coloneqq \int_{\cC} d\br \
        u^*_{n\bk}(\br)
    u_{n'\bk}(\br) = \delta_{nn'}\,,
    \end{align}
    while the normalization volume for Bloch states~$|n\bk\rangle$ is~$\cV$:
\begin{align}
    \langle n\bk|n'\bk\rangle \coloneqq \frac{1}{\cN}\int_{\cV} d\br \
        u^*_{n\bk}(\br)
    u_{n'\bk}(\br) = \delta_{nn'}\,.
    \end{align}
This notation is also used to define an integration of lattice periodic functions over a single unit cell  as
\begin{align}
    \blangle n\bk|f(\br)|n'\bk'\brangle := 
    \int_{\cC}d\br\
    u^*_{n\bk}(\br)
    f(\br)
    u_{n'\bk'}(\br)\,.
\end{align}

In contrast, for expectation values of Bloch states~$|n\bk\rangle$ we integrate over the whole volume~$\cV$ with a normalization~$1/\cN$ stemming from \eqref{eqa3}
\begin{align}
    &\langle n\bk|f(\br)|n'\bk'\rangle  \nonumber
    \\[0.3em]&\overset{\text{\eqref{eqa3}}}{=}
    \frac{1}{\cN}
    \int_{\cV}d\br\
    u^*_{n\bk}(\br)
    e^{-\ci(\bk-\bk')\br}  f(\br)
    u_{n'\bk'}(\br)
    \nonumber
    \\[0.3em]
    &\overset{\text{\eqref{eqa1}}}{=}  
        \frac{1}{\cN} \sum_{\bR}
    \int_{\cC}d\br\
    u^*_{n\bk}(\br)
    e^{-\ci(\bk-\bk')(\br+\bR)}  f(\br+\bR)
    u_{n'\bk'}(\br)\,.
\end{align}
For infinitely extended systems, we have~$\cN{\rightarrow}\,\infty$.

In case we have an operator as the Hamiltonian~$h$ or the density matrix~$\rho$, that are not diagonal in~$\br$, we frequently evaluate matrix elements as follows:
\begin{align}
    \langle& n\bk| h(t) |n'\bk\rangle 
    =
    \iint\limits_{\cV\;\cV} 
    \langle n\bk
    |\br\rangle\langle\br| 
    h(t) 
    |\br'\rangle\langle\br'| 
    n'\bk\rangle \;d\br\, d\br' \nonumber
    \\
    &=\frac{1}{\cN}
    \iint\limits_{\cV\;\cV} 
    \emikr u^*_{n\bk}(\br)
    \langle\br| 
    h(t) 
    |\br'\rangle
        \eikrprime u_{n'\bk}(\br')
    \;d\br\, d\br' \nonumber
    \\
        &\overset{(*)}{=}\frac{1}{\cN}\sum_{\bR}\sum_{\bR'}
    \iint\limits_{\cC\;\cC} 
    \emikrR u^*_{n\bk}(\br)
    \langle\brR| 
    h(t) 
    |\brRprime\rangle\nonumber
    \\&\hspace{10em}
       \times \eikrRprime u^*_{n'\bk}(\br')
    \;d\br\, d\br' \nonumber
        \\
        &\overset{\text{\eqref{eqa3}}}{=}
    \iint\limits_{\cC\;\cC} 
    \blangle n\bk|\br\brangle 
    \Bigg(\frac{1}{\cN} \sum_{\bR}\sum_{\bR'}\emikrR
    \langle\brR| 
    h(t) 
    |\brRprime\rangle\nonumber
    \\&\hspace{10em}
       \times \eikrRprime\Bigg) \blangle\br'|n'\bk\brangle
    \;d\br\, d\br' \nonumber
            \\
  &\overset{(\#)}{=}
  \iint\limits_{\cC\;\cC} 
    \blangle n\bk|\br\brangle 
    \blangle\br| h(\bk;t) |\br'\brangle
    \blangle\br'|n'\bk\brangle
    \;d\br\, d\br' \nonumber
    \\[0.5em]
     &=
     \brank h(\bk;t) \ketnkprime \label{b7}
\end{align}
where we used in~$(*)$ that $u_{n\bk}(\br)$ is lattice-periodic and in the first and last step that the the real-space basis is complete.
In step~$(\#)$, we have defined the operator~$h(\bk;t)$ via its real-space matrix elements as
\begin{align}
    &\blangle\br| h(\bk;t) |\br'\brangle \nonumber
    \\&= 
    \frac{1}{\cN} \sum_{\bR\bR'}\emikrR
    \langle\brR| 
    h(t) 
    |\brRprime\rangle
        \eikrRprime\,.
\end{align}
Bloch states~$|n\bk\rangle$ are eigenstates of the initial, lattice-periodic, time-independent Hamiltonian~$\inh$, 
\begin{align}
    \inh \,|n\bk\rangle = \epsilon_n(\bk)\,|n\bk\rangle\,.
\end{align}
We further have 
\begin{align}
\epsilon_n(\bk)\delta_{nn'}= \langle n\bk| \inh |n'\bk\rangle \overset{\text{\eqref{b7}}}{=}\brank \inh(\bk) \ketnkprime\,.
 \end{align}
After using the completeness~$1=\sum_n\ketnk\brank$ we find the eigenvalue equation for the lattice periodic part
\begin{align}
\inh(\bk) \,\ketnk= 
\epsilon_n(\bk)\ketnk\label{b12}
 \end{align}
 that is used in Eq.~\eqref{e16}.

\section{Matrix elements of local operators \label{a1}} 
We derive an identity relating matrix elements of local operators 
$f(\br)$ in the basis $|n\bk\rangle$ to matrix elements in the basis
$|n\bk\brangle$. 
Employing the basic definitions of periodicity from Appendix~\ref{a0}, 
we have 
\begin{align} 
	&\langle n\bk| f(\br) |n'\bk' \rangle \nonumber\\
	&=
	\frac{1}{\cN}\sum_{\bR} \int_{\cC} d\br\ u^*_{n\bk}(\br)e^{-\ci(\bk-\bk')(\br+\bR)} f(\br+\bR) u_{n'\bk'}(\br)
	\nonumber\\
	&=\frac{1}{\cN}\sum_{\bR} \int_{\cC} d\br\ u^*_{n\bk}(\br)\Big[
	f(\ci\partial_\bq)  e^{-\ci(\bk-\bk'+\bq)(\br+\bR)} \Big]_{\bq=0} u_{n'\bk'}(\br)
	\nonumber\\
	&=\frac{1}{\cN}\Big[f(\ci\partial_\bq)  \sum_{\bR} \blangle 
	n\bk
	|e^{-\ci( \bk-\bk'+\bq)(\br+\bR)}|
	n'\bk'
	\brangle\Big]_{\bq=0}\nonumber\\
	&= \frac{(2\pi)^d}{
	\cV} \Big[f(\ci\partial_\bq)  
	\blangle 
	n\bk
	|e^{-\ci( \bk-\bk'+\bq)\br}\delta(\bk{-}\bk'{+}\bq)|
	n'\bk'
	\brangle\Big]_{\bq=0}\nonumber \\
	&= \frac{(2\pi)^d}{
	\cV} \,
	\blangle 
	n\bk
	|\, f(\ci\partial_\bk) e^{-\ci( \bk-\bk')\br}\delta(\bk{-}\bk')\, |
	n'\bk'
	\brangle \,.
	\label{e:app418} 
\end{align} 
Using identity \eqref{e:app418} and integration by parts, we evaluate $\bk$-sums as follows:  
\begin{align}  
\intbzdkpiprime\, \langle & n\bk|f(\br)|n'\bk'\rangle \psi(\bk') 
\nonumber\\[-0.3em]&= 
\frac{1}{
	\cV}\,\Big[
f(\ci\partial_{\bk'} ) 
 \blangle n\bk|n'\bk'\brangle \psi(\bk') \Big]_{\bk'=
\bk}\,.
\label{eB2} 
\end{align} 
As an application, we consider a Hamiltonian $h(t)$ 
with a vector potential that varies in time and space $\bA(\br,t)$.
The Schr\"odinger dynamics in Bloch-state representation reads 
\be
\ci \partial_t \langle n\bk|\psi\rangle =
\sum_{\un\buk}
 \langle n\bk| h(t)|\un\buk\rangle\langle\un\buk |\psi\rangle\label{ed1}
\ee 
with
\begin{align*} 
\langle n\bk|h(t)|\un\buk\rangle  &\coloneqq \langle n\bk| h(-\ci \nabla - \bA(\br,t))|\un\buk\rangle\,.\nonumber   
\end{align*} 
By virtue of \eqref{eB2}, the rhs matrix element can be rewritten with the consequence that  
\begin{align}
\ci&\partial_t \langle n\bk|\psi\rangle \nonumber\\&= 
\sum_{\un}
\Big[
h(\bk {-} \bA(\ci\partial_{\bk'},t))
\blangle n\bk|\un\bk'\brangle\langle \un\bk'|\psi\rangle\Big]_{\bk=\bk'}.\label{e54} 
\end{align}
As is explicit from this result, the spatial dependency of $\bA(\br)$ mixes neighboring $\bk$-values as a manifestation of the broken translational invariance. 
For a homogeneous $\bA$, however, $\bk$-coupling is absent, as one would expect.

\section{Density matrix in the Coulomb gauge in the co-moving basis} \label{appha}
We derive the expression~\eqref{e40a} for the density matrix in the Coulomb gauge in the co-moving basis~$\ketnkt{=}\ketnksubt$, $\kt{=}\ktA$
\begin{align*}
    \varrho_{nn'}(\bk;t) \coloneqq \blangle n\bk_t|\rho(\bk;t)|n'\bk_t\brangle
\end{align*}
starting from the dynamics~\eqref{e71} in the Coulomb gauge, 
\begin{align*}
    \ci\partial_t\rhoC(\bk{;t}) &= [{\inh}(\bk {-}\bA(t)),\rhoC(\bk{;t})] 
      \,,
\end{align*}
that is projected on the co-moving basis~$\ketnksubt$,
\begin{align}
   \ci\branksubt& (\partial_t\rhoC(\bk;t))\ketnprimeksubt \nonumber
  \\
   &= \branksubt[\inh(\bk {-}\bA(t)),\rhoC(\bk;t)] \ketnprimeksubt\nonumber
      \\&\overset{\text{\eqref{e16}}}{=} \epsilon_{nn'}(\kt)\branksubt \rhoC(\bk;t) \ketnprimeksubt
     \,.\label{eg1}
\end{align}
We are interested in a time derivative of matrix elements instead of matrix elements of the time derivative of operators and therefore state
\begin{align*}
    \branksubt &(\partial_t\rhoC(\bk;t))\ketnprimeksubt =
   \partial_t (\branksubt  \rhoC(\bk;t)\ketnprimeksubt )
   \\&-(\partial_t \branksubt ) \rhoC(\bk;t)\ketnprimeksubt 
   -\branksubt  \rhoC(\bk;t)\partial_t\ketnprimeksubt \,.
\end{align*}
With the resolution of the identity~$1{=}\sum_{\un} \ketunksubt\braunksubt$, $(\partial_t\branksubt)\ketnprimeksubt{=}-\branksubt \partial_t\ketnprimeksubt$ and 
\[
\partial_t\ketnksubt =-\dot\bA(t)\partial_{\kt}\ketnksubt \overset{\text{\eqref{e18a}}}{=} 
\cq\bE(t) \partial_{\kt}\ketnksubt\,,
\]
we arrive at
\begin{align}
\begin{split}
    &\branksubt (\partial_t\rhoC(\bk;t))\ketnprimeksubt =
   \partial_t (\branksubt  \rhoC(\bk;t)\ketnprimeksubt )
   \\[0.8em]&+\cq\bE(t)\sum_{\un} 
   \branksubt(\partial_{\kt}\ketunksubt)
   \braunksubt\rhoC(\bk;t)\ketnprimeksubt 
   \\&
   -\cq\bE(t)\sum_{\un} 
      \branksubt\rhoC(\bk;t)\ketunksubt
   \braunksubt(\partial_{\kt}\ketnprimeksubt )\,.
   \end{split}\label{eg2}
\end{align}
The dipole matrix elements~\eqref{e32}
\begin{align*}
\bfd_{nn'}(\bk_t) =  \cq \bracom\ci\partial_{\bk_t}\ketcomprime
\end{align*}
together with Eqs.~\eqref{eg1} and~\eqref{eg2} lead to
\begin{align}
    \big(\ci\partial_t-&\epsilon_{nn'}(\kt) \big) \branksubt  \rhoC(\bk;t)\ketnprimeksubt  =\nonumber
    \\[0.5em]&
\bE(t)\sum_{\un}\Big[
\branksubt\rhoC(\bk;t)\ketunksubt \bfd_{\un n'}(\kt) \nonumber
\\[-0.7em]&
\hspace{3em}-
\bfd_{n\un}(\kt)\braunksubt\rhoC(\bk;t)\ketnprimeksubt \Big]\,.\label{eg3}
\end{align}
Eq.~\eqref{eg3} is identical to Eq.~\eqref{e33a} and we conclude Eq.~\eqref{e40a},
\begin{align}
    \branksubt\rhoC(\bk;t)\ketnprimeksubt = \varrho_{nn'}(\bk;t)\label{eg4}\,.
\end{align}

\section{Illustrating the boost operator and proof of Eq.~\eqref{e48d}}\label{appga}
The boost operator has been defined in the main text in Eq.~\eqref{e41c} as
\begin{align*}
    \mfB(t) = \mfBdef  \,.
\end{align*}
We consider this operator as a successive, time-ordered infinitesimal shifting,
\begin{align}
    \mfB(t)\cong \prod_{t'=-\infty}^t \Big(1+dt'\,\dot\bk(t')\partial_\bk\Big)\,.
\end{align}
We use $\dot\bk(t)\eqt\partial_t (\bk{-}\bA(t))\eqt{-}\dot\bA(t)$ for a homogeneous electric field and Taylor expansion $f(\bk{-}dt\,\dot\bA(t))\eqt(1{-}dt\dot\bA\partial_\bk)f(\bk)$  
to show 
\begin{align*}
 \mfB(t) f(\bk) &\cong 
  \prod_{t'=-\infty}^t \Big(1-dt'\,\dot\bA(t')\partial_\bk\Big)\;f(\bk)
  \\[0.3em]
  &= f\big(\bk-\smallint  dt' \dot\bA(t')\big) = f(\ktA)\,.
\end{align*}

Next, we prove Eq.~\eqref{e48d}, 
\begin{align*}
    \varrho_{nn'}(\bk;t) = \rhoDnnprime(\bk{-}\bA(t);t)\,. 
\end{align*}
We start by specifying the inverse of~$\mfB$,
\begin{align}
    \mfB^{-1}(t) = \mfBdefinv
\end{align}
and stating
\begin{align}
    \frac{d}{dt}\mfB^{-1}(t) = -\cq\bE(t)\partial_\bk \mfBdef
    \label{eh3}
\end{align}
where we have used $\dot\bk(t)\eqt\partial_t (\bk{-}\bA(t))\eqt{-}\dot\bA(t)\eqt\cq\bE(t)$ for a homogeneous electric field. 
We apply~$\mfB^{-1}(t)$ to the left of the $\varrho$ dynamics, Eq.~\eqref{e33a}, and obtain (suppressing the time dependence of $\mfB$)
\begin{align*} 
  &\ci \mfB^{-1}\frac{d}{dt}\varrho_{nn'}(\bk;t)
- \epsilon_{nn'}(\bk)\mfB^{-1}\varrho_{nn'}(\bk;t)
 =\\[0.3em]
&\bE(t) 
\sum_{\un} 
\left(\mfB^{-1}\varrho_{n\un}(\bk;t)\right)\bfd_{\un n'}(\bk)
-\,\bfd_{n\un}(\bk )\mfB^{-1}\varrho_{\un n'}(\bk;t)\,.
\end{align*} 
We insert Eq.~\eqref{eh3} and obtain
\begin{align} 
\begin{split}
  \Big(&\ci\frac{d}{dt} +\ci\cq\bE(t)\partial_\bk
-\epsilon_{nn'}(\bk)\Big)\mfB^{-1}\varrho_{nn'}(\bk;t)
 =\\[0.3em]
&\bE(t) 
\sum_{\un} 
\left(\mfB^{-1}\varrho_{n\un}(\bk;t)\right)\bfd_{\un n'}(\bk)
-\,\bfd_{n\un}(\bk )\mfB^{-1}\varrho_{\un n'}(\bk;t)\,.\label{eh4}
\end{split}
\end{align}
The {\eom} for $\mfB^{-1}\varrho_{nn'}(\bk;t)$ in Eq.~\eqref{eh4} is identical to the {\eom} of $\rhoD_{nn'}(\bk;t)$ in Eq.~\eqref{e45} and we conclude
\begin{align}
    \mfB^{-1}(t)\varrho_{nn'}(\bk;t) = \rhoD_{nn'}(\bk;t)\,.
\end{align}
Eq.~\eqref{e48d} follows.

\section{Dynamical polarization~$\bP$ in $\bk-$dependent basis}\label{aH}
In the main text, we derive the dynamical polarization in a Bloch basis with $\bk$-independent lattice-periodic part, see Eq.~\eqref{e75b}.
In this Appendix, we compute the polarization as expectation value of the dipole operator $\cq\br$  in the stationary Bloch basis~$|n\bk\rangle$ from \eqref{e15} as~\cite{Schaefer2002}
\bea
{\bf P}(t) &=& \frac{1}{\cV}\sum_{\alpha,\beta} \langle \alpha | \cq\br|\beta \rangle \rho_{\beta\alpha}(t) 
\nonumber\\
&=&\frac{1}{\cV}\sum_{nn'}\sum_{\bk\bk'} \langle n\bk|  \cq\br|n'\bk' \rangle \rho_{n'n}(\bk'\bk;t)\,,
\label{ae39a}
\eea
with 
$\cV$ denoting the normalization volume.
We keep the full $\bk$-dependence of
\[
\rho_{nn'}(\bk\bk';t) = \langle n\bk|\rho(t)|n'\bk'\rangle
\]
in \eqref{ae39a} to properly account for $\bk$-derivatives of dipole matrix elements later on.
%

We evaluate the dipole matrix element $\langle n\bk| \br|n'\bk' \rangle $ appearing in the polarization \eqref{e39a} adopting \eqref{e:app418} as
\begin{align} 
	\langle n\bk| \br |n'\bk' \rangle 
	=\frac{(2\pi)^d}{\cV} 
	\blangle n\bk|\left[\ci\partial_\bk e^{-\ci( \bk-\bk')\br}\delta(\bk{-}\bk')\right]|n'\bk'\brangle\,.\label{ae41a}
\end{align} 
With \eqref{e41a} and results from Appendix~\ref{a0}, we obtain
\begin{align}
\bP(t) 
&= \cq \sum_{nn'}\intbzdk\intbzdkpiprime\ \rho_{n'n}(\bk'\bk;t) \nonumber
\\
&\hspace{3em} 	\times\blangle n\bk|\left[\ci\partial_\bk e^{-\ci( \bk-\bk')\br}\delta(\bk{-}\bk')\right]|n'\bk'\brangle\,.\nonumber
\\[0.6em]
&= \ci \cq  \sum_{nn'} 
\intbz \frac{d\bk}{(2\pi)^d}\ 
\Big( 
\blangle n\bk|\partial_{\bk}|n'\bk\brangle
\rho_{n'n}(\bk\bk;t) \nonumber
\\[-0.3em]
& \hspace{4em}+  \blangle n\bk|n'\bk\brangle\left.\frac{\partial\rho_{n'n}(\bk^\prime\bk;t)}{\partial 
\bk^\prime}\right|_{\bk'\to\bk}
\Big)
\label{ae42}
\end{align} 
where integration by parts has been used to arrive at the last equation.
We define 
\be 
\brho_{n'n}(\bk;t) \coloneqq \langle c^\dagger_{n\bk}\partial_{\bk} c_{n'\bk}\rangle =
\left.\frac{\partial\rho_{n'n}(\bk^\prime\bk;t)}{\partial 
\bk^\prime}\right|_{\bk'\to\bk}\label{ae41}
\ee 
so that 
\begin{align}
   \bP
=
 \sum_{n n'} \intbzdkpi\ \Big( {\bf d}_{nn'}(\bk)\ \rho_{n'n}(\bk;t)
+ \ci \cq \delta_{nn'} \brho_{n'n}(\bk;t) \Big) 
\label{ae39} 
\end{align}
recalling 
\begin{align}
    \bfd_{nn'}(\bk) = 
\blangle n\bk|\ci\cq\partial_\bk|n'\bk\brangle \label{ae53c}
\end{align}
and abbreviating 
\begin{align}
    \rho_{nn'}(\bk;t)&\coloneqq\rho_{nn'}(\bk\bk;t)  = \langle n\bk|\rho(t)|n'\bk\rangle \nonumber
    \\&
    \overset{\text{\eqref{b7}}}{=}\brank \rho(\bk;t)\ketnkprime
    \,.\label{ae53a}
\end{align} 

The time derivative~$\dtP$ of the polarization is needed for evaluating the emission~\eqref{e39b} and is given by 
\begin{align}
   \dtP
&=
 \sum_{n n'} \intbzdkpi\ \Big( {\bf d}_{nn'}(\bk)\ \dot\rho_{n'n}(\bk;t)
+ \ci \cq \delta_{nn'} \dot\brho_{n'n}(\bk;t)\Big) \label{ae43}
\nonumber\\
&= 
\intbzdkpi\ \Trn \Big(\mathbf{d}(\bk)\dot\rho(\bk;t) + \ci\cq \dot\brho(\bk;t)\Big)
\end{align}
 After inserting the EoM  \eqref{e71}, $\ci \dot \rho(\bk;t) = [h(\bk;t), \rho(\bk;t)] $ in the Coulomb gauge  in the rhs of~\eqref{ae41} the second term of \eqref{ae43} contributes with 
\begin{align}
\ci \sum_{n} \dot \brho_{nn}(\bk;t)
&= \sum_{nn'}(\partial_\bk h_{nn'}(\bk,t))\rho_{n'n}(\bk;t) 
\label{ae53} 
\end{align}
so that the coupling to $\brho(\bk;t)$ drops out due to cyclic invariance of the trace. 
Recalling the EoM for $\rho(t)$  we arrive at:
\begin{align}
\partial_t {\bf P} 
&=
 \sum_{n n'} \intbzdkpi\ 
\Big( [-\ci{\bf d}(\bk),h(\bk;t)]_{nn'} 
\nonumber\\[-0.4em] &\hspace{6em}
+ 
\cq \partial_\bk h_{nn'}(\bk,t) \Big)\ \rho_{n'n}(\bk;t)  
\nonumber
\\[0.6em] 
&= 
\cq  \sum_{n n'}  \intbzdkpi\ 
\blangle n\bk|\frac{\partial h(\bk;t)}{\partial \bk}|n'\bk\brangle \,\rho_{n'n}(\bk;t) 
\nonumber\\&
= \cq \intbzdkpi \ \text{Tr}_n \left[
\frac{\partial h(\bk;t)}{\partial \bk} \,\rho(\bk;t)\right]
\label{ae40}
\end{align}
The last line uses 
\begin{align}
&\partial_\bk h_{nn'}(\bk,t) 
= \partial_\bk  \blangle n\bk|h(\bk;t)|n'\bk\brangle 
\nonumber
\\
&= \blangle n\bk|h\partial_\bk|n'\bk\brangle + 
\blangle n\bk | \frac{\partial h}{\partial \bk} |n'\bk\brangle + 
\blangle n\bk|\partial_\bk^\dagger h |n'\bk\brangle \,.
\label{ae49a} 
\end{align} 
With Eq.~\eqref{ae40}, we arrive at the same result as in Eq.~\eqref{e42c} that has obtained using the Bloch basis with $\bk$-independent lattice-periodic part.

\section{Proof of Eq.~\eqref{e86}}
\label{apph}
For the proof of Eq.~\eqref{e86},
\begin{align*}
   \blangle n\bk|\frac{\partial \inh(\bk)}{\partial\bk}|n'\bk\brangle
   = \delta_{nn'}\frac{\partial \epsilon_n(\bk)}{\partial \bk} 
   + \frac{\ci}{\cq}\, \epsilon_{nn'}(\bk) \bfd_{nn'}(\bk)\,, 
\end{align*}
we execute
\begin{align*}
   &\delta_{nn'}\partial_\bk\epsilon_n(\bk) =  \partial_\bk \brank\inh(\bk)\ketnkprime
    \\
    &= \blangle n\bk|\inh\partial_\bk|n'\bk\brangle + 
\blangle n\bk | \frac{\partial \inh}{\partial \bk} |n'\bk\brangle + 
\blangle n\bk|\partial_\bk^\dagger \inh |n'\bk\brangle 
\\
    &= \epsilon_n(\bk)\blangle n\bk| \partial_\bk|n'\bk\brangle + 
\blangle n\bk | \frac{\partial \inh}{\partial \bk} |n'\bk\brangle - 
\epsilon_{n'}(\bk)\blangle n\bk|\partial_\bk |n'\bk\brangle
\\
&\overset{\text{\eqref{e32}}}{=}
-\epsilon_n(\bk)\frac{\ci}{\cq}\dnnprimek  + 
\blangle n\bk | \frac{\partial \inh}{\partial \bk} |n'\bk\brangle + 
\epsilon_{n'}(\bk)
\frac{\ci}{\cq}\dnnprimek\,,
\end{align*}
where we have used $\brank\partial_\bk^\dagger\ketnkprime\eqt{-}\brank \partial_\bk\ketnkprime$.
Eq.~\eqref{e86} follows with $\epsilon_{n'n}(\bk)\eqt\epsilon_{n'}(\bk){-}\epsilon_{n}(\bk)$.

\section{Derivation of the anomalous velocity and the conductivity tensor in Eq.~\eqref{e59}/\eqref{e96}}
\label{sec:appg}
For the proof of Eq.~\eqref{e59}/\eqref{e96}, 
 \begin{align*}
     \bJ(t) = \jintra  + \dtPinter + 
     \sigmainter \bE(t) +\Jdamp \,,
\end{align*}
we start from Eq.~\eqref{e87}/\eqref{e75}
\begin{align} 
 \bj(t) &= \jintratilde +\jintertilde \nonumber
 \\
 &= \jintratilde 
 -\ci 
  \sum_{n\neq n'} \intbzdkpi\
  \mathbf{d}_{nn'}(\bk)\ \epsilon_{n'n}(\bk)
  \rhoDnprimen(\bk;t)\,.\label{eg1a}
 \end{align}
Replacing~$\epsilon_{n'n}(\bk)\rhoD_{n'n}(\bk;t)$ in Eq.~\eqref{eg1a} by the \eom~\eqref{e99} and using the definitions~\eqref{e69},~\eqref{e95a}  leads to 
  \begin{align} 
 \bj&(t) = \jintratilde +  \dtPinter + \Jdamp
  \nonumber
  \\
  &+
  \sum_{n\neq n'} \intbzdkpi\
  \mathbf{d}_{nn'}(\bk)\
  \cq \bE(t)\partial_\bk
  \rhoDnprimen(\bk;t)\,
  \nonumber\\
  & -\ci\sum_{n\neq n'} \intbzdkpi\
  \mathbf{d}_{nn'}(\bk)\
  [\bE(t)\mathbf{d}(\bk),\rhoD(\bk;t)]_{n'n}
  \,.
  \label{eg2a}
 \end{align}
We focus on the $i$th component of the last term. 
Suppressing the time and $\bk$-dependence, we have:
{\allowdisplaybreaks
\begin{align}
   \sum_{n\neq n'}& d^{(i)}_{nn'}\
  [\bE\cdot\mathbf{d},\rhoD]_{n'n}\nonumber =  \\
  =& \sum_j E^{(j)}\hspace{-0.4em}\sum_{\un,n\neq n'} \hspace{-0.2em}
  d^{(i)}_{nn'}
  \left(
  d^{(j)}_{n'\un}\rhoD_{\un n}- \rhoD_{n'\un}d^{(j)}_{\un n}
  \right)\nonumber
  \\
  =&\sum_j E^{(j)} \Big[
  \sum_{n, n', \un}  
  d^{(i)}_{nn'}
  \left(
  d^{(j)}_{n'\un}\rhoD_{\un n}- \rhoD_{n'\un}d^{(j)}_{\un n}
  \right)
  \nonumber
  \\
  &\hspace{3em}  -\sum_{n\un} d_{nn}^{(i)}
  \left( 
  d_{n\un}^{(j)}\rhoD_{\un n} - 
  \rhoD_{n\un}d^{(j)}_{\un n}
  \right)
  \Big]
  \nonumber
   \\
  =&
  \sum_j E^{(j)} 
\sum_{nn'} 
\left(
d^{(j)}_{nn'}d^{(i)}_{n'n} - 
d^{(i)}_{nn'}d^{(j)}_{n'n}
\right)
\rhoD_{nn}
\nonumber
\\
&+
  \sum_j E^{(j)} 
  \sum_{n\neq \un}  
  \Big[
  \sum_{n'}
  \left(
  d^{(i)}_{nn'}d^{(j)}_{n'\un} - d^{(j)}_{nn'}d^{(i)}_{n'\un}
  \right)\nonumber
  \\
  &\hspace{5.5em}
  -d^{(j)}_{n\un}
  \left(
  d^{(i)}_{nn}-d^{(i)}_{\un\un}
  \right)
  \Big]
  \rhoD_{\un n}\nonumber
\\[0.3em]
=& \ci\cq\sum_n \,
[\bE(t) \times \bomega_n(\bk)]^{(i)}
\,\rhoD_{nn}
\nonumber
\\
&+
\sum_j E^{(j)} \sum_{n\neq \un}
\Big[
d^{(j)}_{n\un}
\left(
d^{(i)}_{\un\un} -
d^{(i)}_{nn}
\right)
+
[d^{(i)},d^{(j)}]_{n\un}
\Big]
\rhoD_{\un n}
\label{eg3a}
\end{align}

In the last step, we have identified the Berry curvature~$\bomega_n(\bk)$  by using the sum rule\cite{Aversa1995}  
\begin{align*}
   [& d^{(i)},d^{(j)}]_{nn} = \sum_{n'}
  \left(
  d^{(i)}_{nn'}d^{(j)}_{n'n} - d^{(j)}_{nn'}d^{(i)}_{n'n}
  \right)
  \\
  &\overset{\eqref{e32}}{=} 
  -\cq^2
  \sum_{n'} \Big(
 \brank \partial_{k_i} \ketnkprime
 \brankprime \partial_{k_j}\ketnk
 \\[-0.6em]&\hspace{6em}-
  \brank \partial_{k_j} \ketnkprime
 \brankprime \partial_{k_i}\ketnk
 \Big)
  \\
    &=
  -\cq^2
  \sum_{n'} \Big(-
 \brank \partial_{k_i}^\dagger \ketnkprime
 \brankprime \partial_{k_j}\ketnk
 \\[-0.6em]&\hspace{6em}+
  \brank \partial_{k_j}^\dagger \ketnkprime
 \brankprime \partial_{k_i}\ketnk
 \Big)
  \\
      &=
  -\cq^2
    \left(-
 \brank \partial_{k_i}^\dagger  
 \partial_{k_j}\ketnk
+
  \brank \partial_{k_j}^\dagger 
  \partial_{k_i}\ketnk
 \right)
  \\
          &=
  -\cq^2
    \Big(-\partial_{k_i} 
 \brank   
 \partial_{k_j}\ketnk
 +
  \brank \partial_{k_i} 
 \partial_{k_j}\ketnk
 \\&\hspace{4em}
 +
 \partial_{k_j}  \brank 
  \partial_{k_i}\ketnk
  -
  \brank \partial_{k_j} 
  \partial_{k_i}\ketnk
 \Big)
  \\[0.3em]
 & = \ci\cq 
  \left(
  \partial_{k_j}d^{(i)}_{nn}-
  \partial_{k_i}d^{(j)}_{nn}
  \right)
  \\[0.3em]
 & =
 \ci\cq \sum_{ab}
 \left(
 \delta_{aj}\delta_{bi}-
 \delta_{bj}\delta_{ai}
 \right)
 \partial_{k_a}d^{(b)}_{nn}
  \\[0.3em]
  & =
 \ci\cq \sum_{abc}
 \epsilon_{abc}\,
 \epsilon_{cij}\,
 \partial_{k_a}d^{(b)}_{nn}
=
 \ci\cq \sum_{c}
 \epsilon_{ijc}\,
 \Omega_n^{(c)}
\end{align*}
where in the last line, we have used the definition of the Berry curvature from Eq.~\eqref{e72a} written with the Levi-Civita tensor~$\epsilon_{abc}$.
We will also need a similar sum rule\cite{Aversa1995} that is
\begin{align}
    \ci 
    [
    d^{(j)} , d^{(i)}
    ]_{nn'}
    =
    \cq\left( 
    \partial_{k_j}d^{(i)}_{nn'}
    -
    \partial_{k_i}d^{(j)}_{nn'}
    \right)\,.\label{eg4a}
\end{align}
Using integration by parts in the fourth term of Eq.~\eqref{eg2a} and inserting \eqref{eg3a} into \eqref{eg2a}, we obtain
  \begin{align} 
 \bj&(t) =\, \jintratilde +  \dtPinter +\Jdamp
  \nonumber
  \\
  &-
  \cq
  \sum_{j}E^{(j)}(t)
  \sum_{n\neq n'}
  \intbzdkpi\
  \left(\partial_{k_j}\bfd_{nn'}(\bk)\right)
  \rhoDnprimen(\bk;t)\,
  \nonumber\\
  & + \cq \sum_n\intbzdkpi\;\bE(t) \times \bomega_n(\bk)\,\rhoD_{nn}(\bk;t)
  \nonumber\\
  & -\ci\sum_j E^{(j)}(t)\sum_{n\neq n'} \intbzdkpi\
  \Big[
d^{(j)}_{nn'}
\left(
\bfd_{n'n'}(\bk) -
\bfd_{nn}(\bk)
\right)\nonumber
\\
&\hspace{9em}+
[\bfd,d^{(j)}(\bk)]_{nn'}
\Big]
\rhoD_{n'n}(\bk;t)
\nonumber
\\
\overset{\eqref{eg4a}}{=}& \;\,\jintratilde +  \dtPinter+\Jdamp
  \nonumber
  \\
  & + \cq \sum_n\intbzdkpi\;\bE(t) \times \bomega_n(\bk)\,\rhoD_{nn}(\bk;t)
  \nonumber\\
  & +\sum_j E^{(j)}(t)\sum_{n\neq n'} \intbzdkpi\
 \ci \Big[
d^{(j)}_{nn'}
\left(
\bfd_{nn}(\bk) -
\bfd_{n'n'}(\bk)
\right)\nonumber
\\
&\hspace{9em}-
\left(\partial_{\bk}d_{nn'}^{(j)}(\bk)\right)
\Big]
\rhoD_{n'n}(\bk;t)\,.
  \label{eg5a}
 \end{align}
Eqs.~\eqref{e72}, \eqref{e70}, \eqref{e59} and~\eqref{e96} follow.
}

\begin{figure*}[t]
    \centering
\includegraphics[width=\textwidth]{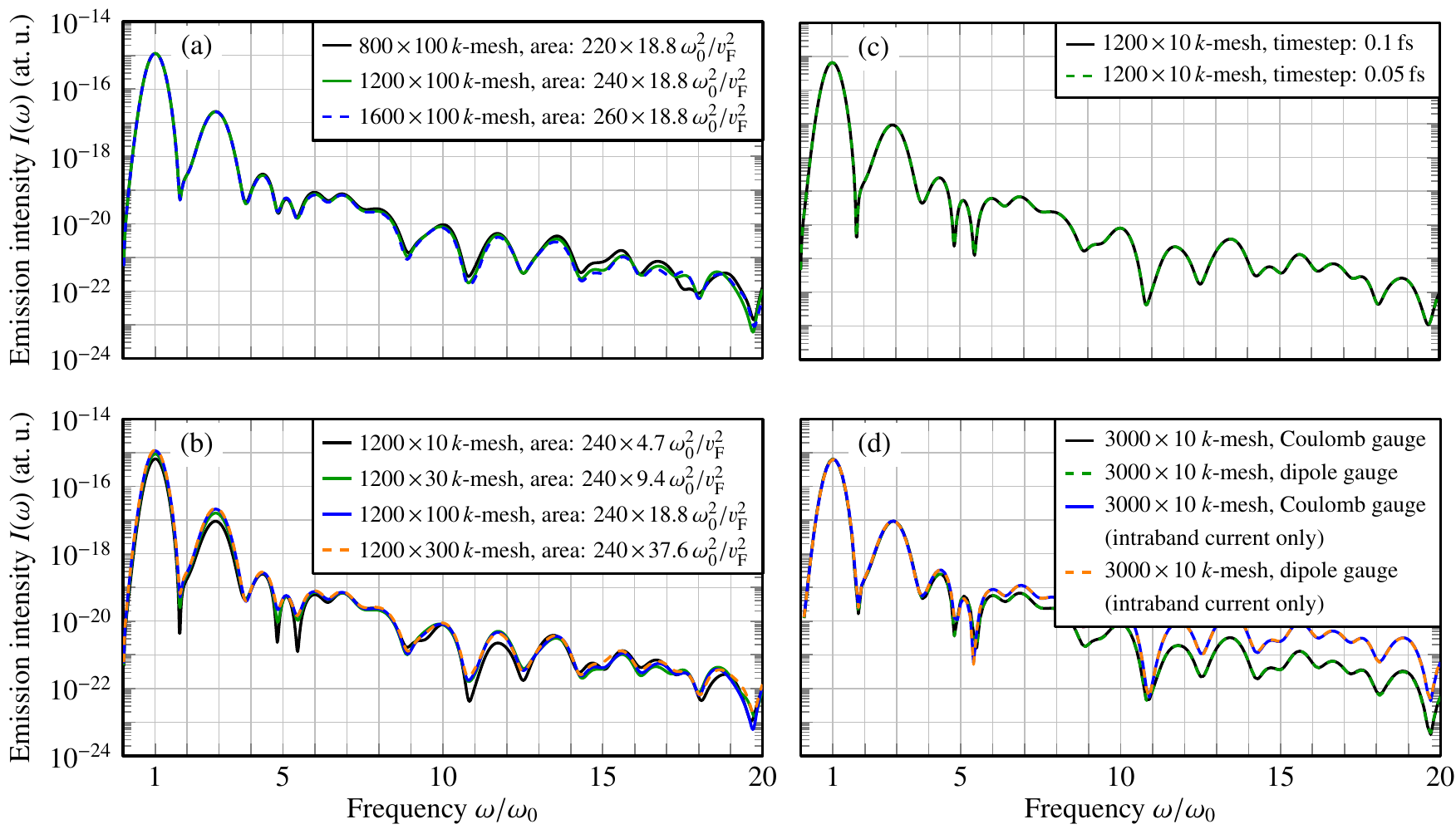}
    \caption{Frequency-dependent emission intensity from dynamics in a Dirac-cone bandstructure 
    when increasing the $k$-point density and the edge length of the rectangle
    (a)  in $k_x$ direction and (b) in $k_y$ direction.  
    Convergence is found for a rectangle of size 1500\,$\times$\,240\,$\omega_0^2/\vF^2$ and a 1200\,$\times$\,100 $k$-mesh resolution.
    (c) Emission intensity computed with two different timesteps (0.1\,fs and 0.05\,fs) when propagating the {\eom} finding excellent agreement between both traces.
    (d) Emission intensity computed from the Coulomb and dipole gauge for a 3000\,$\times$\,10 $k$-mesh finding good agreement. Shown are the emission from the total current and from the intraband current.
    }
    \label{fig:DiracwithoutB}
\end{figure*}
\begin{figure}[b]
    \centering
\includegraphics[width=8.5cm]{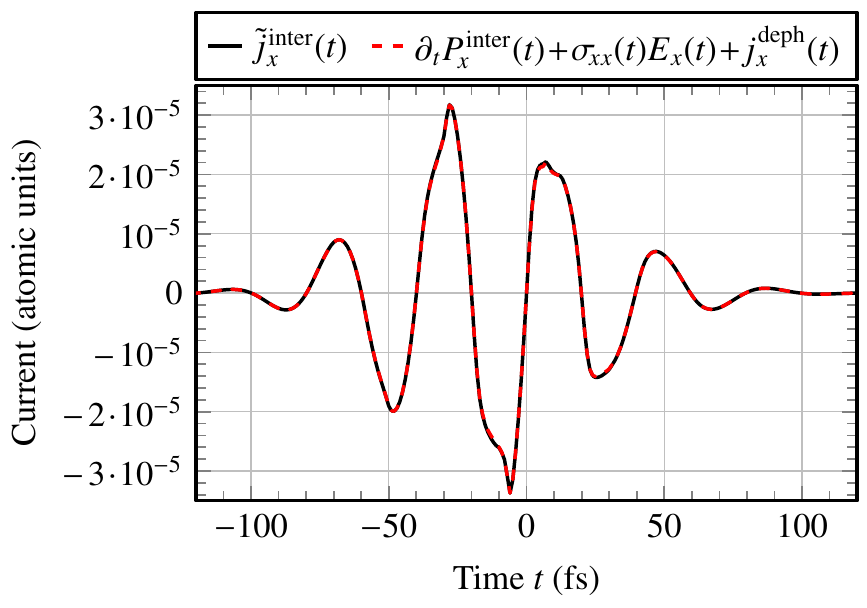}
    \caption{The component of the time-dependent interband current~$\jintertildex$  directed along the electric field computed from Eq.~\eqref{e69a} 
    (black) and  from Eq.~\eqref{e96} (red dashed).
    }
    \label{fig:currenttotalinta}
\end{figure}

\section{Convergence tests}\label{sec:apph}
{\bf k-meshes.} The singular nature of the Dirac-Hamiltonian manifests in divergences near the $\Gamma$-point, e.g. in the dipole-matrix elements. Therefore, particular care has to be taken when discretizing the $k$-space. We here investigate the convergence of the $k$-point integration of Eq.~\eqref{e95} in Fig.~\ref{fig:DiracwithoutB}\,(a) and\,(b).
As $k$-point mesh, we choose a  $\Gamma$-centered Monkhorst-Pack mesh~\cite{Monkhorst1976} that is confined by a rectangle.
As shown in Fig.~\ref{fig:DiracwithoutB}\,(a), the emission intensity converges when increasing the size of the $k$-mesh and the density of $k$-points in direction of the $E$-field ($x$-direction).
Convergence is found for 1200 $k$-points and a length~240\,$\omega_0/\vF$ in $k_x$-direction.

From Fig.~\ref{fig:DiracwithoutB}\,(b), we observe that the emission intensity converges when increasing the length and density of the $k$-mesh orthogonal to the electric driving field ($k_y$-direction).
Here, convergence is found for 100 $k$-points and a length of 18.8\,$\omega_0/\vF$ in $k_y$-direction.
We  are left to choose a rectangular 1200\,$\times$\,100 $k$-mesh with size~240\,$\times$\,18.8\,$\omega_0^2/\vF^2$ for all $k$-integrations from the main text.
As maximum time step, we choose 0.1\,fs.
Decreasing the time step to 0.05\,fs hardly changes the emission curves, see Fig.~\ref{fig:DiracwithoutB}\,(c).  

{\bf Gauge independence.} As an extra numerical test proving the
equivalence of gauges, we compute the current in dipole gauge from Eq.~\eqref{e85} and in the Coulomb gauge from Eq.~\eqref{e95}.
The high-harmonics spectrum follows from Eq.~\eqref{e39b}. 
Fig.~\ref{fig:DiracwithoutB}\,(d) displays our results:
Two emission curves are shown for currents computed in the Coulomb and dipole gauge that lie on top of each other demonstrating the expected equivalence of gauges.

{\bf Current formul{\ae}.} 
A key result of our paper is the decomposition formula 
\eqref{e59} and \eqref{e96}. 
We demonstrate in  
Fig.~\ref{fig:currenttotalinta}
its equivalence with respect to the interband current
to the pre-decomposed expression \eqref{e69a} for the Dirac model
and a current component directed along the electric field, say $x$-direction. 
In this simple setup, 
the anomalous contribution to the velocity~$(\bE(t)\timest\bomega_n(\bk))_x$ vanishes; we have for the $x$-component of the interband current 
\begin{align}
    \jintertildex = \dtPinterx\pt\sigma_{xx}(t)E(t)+\jdampx\,.\label{eh2}
\end{align}
As seen from Fig. 
\ref{fig:currenttotalinta} 
the results obtained from both calculation methods indeed agree, as they should.

\section{Emission from semiconductor Hamiltonian}\label{sec:appsemic}

In Sec.~\ref{sec5}, we have applied the {\sbe} formalism to Dirac fermions. 
For a comparison to previous {\sbe} studies in semiconductors~\cite{\refsemissionformula}, we here investigate as a toy model for a semiconductor a one-dimensional two-band Hamiltonian
\begin{align}
\inh(k) = t(k)\sigma_x + \Delta\sigma_z\,,
\hspace{1em}
t(k) \coloneqq t(1 + \cos (ka))\label{ei1}
\end{align}
for $k\,{\in}\,(-\pi/a,\pi/a]$ that has a semiconducting spectrum
\begin{align}
    \epsilon_c(k)=-\epsilon_v(k) = \sqrt{\Delta^2+t^2(k)}
\end{align}
with minimal gap~$2\,\Delta$.
As parameters, we choose $t\eqt3\,\text{eV}$, $\Delta\eqt1.5\,\text{eV}, a\eqt3$\,{\AA} to mimic a generic semiconductor.
As one-dimensional driving field, we choose a Gaussian pulse as in Eq.~\eqref{e91},
\begin{align}
     E(t)  = E\,  \sin(\omega_0 t)\, \exp\left(-\frac{t^2}{\sigma^2}\right)\,,\label{ei3}
\end{align}
with $\omega_0\eqt2\pi\,{\cdot}\,90\,\text{THz}$, $E \eqt 10\,\text{MV/cm}$ and~$\sigma\eqt25$\,fs.
\begin{figure}[t!]
    \centering
    \includegraphics[width=8.5cm]{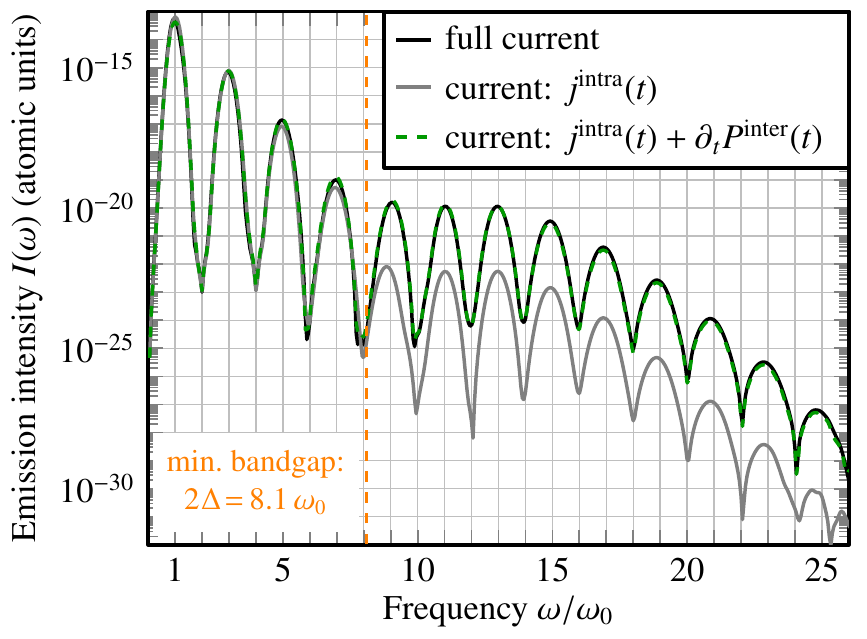}
    \caption{Emission intensity~$I(\omega)$, Eq.~\eqref{e39b}, computed from different currents for a toy model of a one-dimensional semiconductor: 
    black: full current using Eq.~\eqref{e85}, 
    gray dashed: intraband current~$\jintra$ from Eq.~\eqref{e72},
    dashed blue: sum of intraband current~$\jintra$  and polarization related current~$\dtPinter$ from Eq.~\eqref{e69}.
    }
    \label{fig:emissionsemic}
\end{figure}

Solving the {\sbe}~\eqref{e99} in the dipole gauge with $T_2\eqt 1$\,fs  and computing currents from Eqs.~\eqref{e85}, \eqref{e59} and \eqref{e96} results in a frequency-dependent emission~$I(\omega)$ [Eq.~\eqref{e39b}] that is shown in Fig.~\ref{fig:emissionsemic}.
Only odd harmonics appear trivially, due to the inversion-symmetric Hamiltonian
\eqref{ei1}, $\inh(k)\eqt\inh(-k)$. 
We observe phenomenology similar to Ref.~\onlinecite{Vampa2014}: a perturbative regime exists for the first to seventh harmonic that correspond to frequencies~$\omega$ below the minimal bandgap~$2\,\Delta\apt 8\,\omega_0$.
A plateau follows up to 15th harmonic order followed by an exponential decay for harmonics exceeding~$15\,\omega_0$. 

Also in line with Ref.~\onlinecite{Vampa2014}, we observe in Fig.~\ref{fig:emissionsemic} that the emission from the full current exceeds the emission from the intraband current by orders of magnitude for frequencies above the minimum bandgap. 
In addition, the emission from the sum of intraband current and time-derivative of the interband-polarization is very close to the emission from the full current in clear contrast to Dirac fermions shown in Fig.~\ref{fig:emission}. 
Thus, we confirm that for this application the corrections to the traditional approxmation for the total current, $\jintrascalar\pt\dtPinterscalar$, are indeed small.  

\bibliography{Literature}

\end{document}